\documentclass[10pt,twocolumn,twoside]{IEEEtran}
\usepackage{enumitem}
\usepackage{amsmath,mathrsfs,amssymb}
\usepackage{url}
\usepackage{setspace}
\usepackage[utf8]{inputenc}
\usepackage{mathbbol}
\usepackage{xr}
\usepackage{subcaption}

\usepackage{preamble}
\let\chapter\section
\usepackage[ruled,vlined]{algorithm2e}

\usepackage[style=ieee, citestyle=numeric-comp, natbib=false, sorting=none, 
uniquename=false, hyperref=false, backend=bibtex, bibencoding=auto, 
minnames=1,defernumbers=true, maxcitenames=2, maxbibnames=6]{biblatex}

\usepackage{balance}

\AtEveryBibitem{\clearfield{doi}}
\AtEveryBibitem{\clearfield{issn}}
\AtEveryBibitem{\clearfield{isbn}}

\usepackage[final,factor=1100,stretch=10,shrink=10]{microtype}

\newtheorem{thm}{Theorem}
\newtheorem{lem}{Lemma}

\newtheorem{cor}{Corollary}
\newtheorem{rem}{Remark}

\newtheorem{definition}{Definition}

\title{Projected Nesterov's Proximal-Gradient Algorithm for \kern-1pt 
Sparse \kern-1pt Signal  \kern-1pt Reconstruction \kern-1pt with a Convex 
Constraint}

\author{
  \IEEEauthorblockA{Renliang Gu and Aleksandar Dogandžić}
  \thanks{The authors are with the Department of Electrical and Computer 
    Engineering, Iowa State University, Ames, IA 50011 USA (e-mail:
    \texttt{\{\mbox{renliang},ald\}@iastate.edu}).
    This work was supported by the U.S.\ National Science Foundation under 
    Grant CCF-1421480.
  }
}

\newcommand*{\putNum}[4]{
  \put(0,#2){
    \color{#4}\makebox[0pt][r]{\small RSE=\SI{#1}{\percent}\rule{#3mm}{0pt}}
  }
}

\usepackage{todonotes}

\begin{document}

\maketitle

\begin{abstract}
   We develop a \gls{PNPG} approach for sparse signal reconstruction that 
   combines adaptive step size with Nesterov's momentum acceleration.  The 
   objective function that we wish to minimize is the sum of a convex 
   differentiable data-fidelity (\gls{NLL}) term and a convex 
   regularization term.  We apply sparse signal regularization where the 
   signal belongs to a closed convex set within the closure of the domain 
   of the \gls{NLL};
   the convex-set constraint facilitates flexible \gls{NLL} domains and 
   accurate signal recovery.  Signal sparsity is imposed using the 
   $\bm{\ell}_1$-norm penalty on the signal's linear transform coefficients 
   or gradient map, respectively.  The \gls{PNPG} approach employs 
   \emph{projected} Nesterov's acceleration step with restart and an inner 
   iteration to compute the proximal mapping. We propose an \emph{adaptive} 
   step-size selection scheme to obtain a good local majorizing function of 
   the \gls{NLL} and reduce the time spent backtracking.
   Thanks to step-size adaptation, \gls{PNPG} \emph{does not} require 
   Lipschitz continuity of the gradient of the \gls{NLL}.  We present an 
   integrated derivation of the momentum acceleration and its 
   $\bm{\mathcal{O}(k^{-2})}$ convergence-rate and iterate convergence 
   proofs, which account for adaptive step-size selection, inexactness of 
   the iterative proximal mapping, and the convex-set constraint.   The 
   tuning of \gls{PNPG} is largely application-independent. 
   %
   %
   %
  Tomographic and compressed-sensing reconstruction experiments with 
  Poisson generalized linear and Gaussian linear measurement models 
  demonstrate the performance of the proposed approach.
\end{abstract}

\glsresetall
\glsunset{NPGS}
\glsunset{CPU}

\section{Introduction}
\label{sec:introduction}


Most natural signals are well described by only a few significant 
coefficients in an appropriate transform domain, with the number of 
significant coefficients much smaller than the signal size.  Therefore, for 
a vector $\bx \in \mathamsbb{R}^{p \times 1}$ that represents the signal 
and an appropriate \emph{sparsifying transform} $\bpsi(\cdot): 
\mathamsbb{R}^p \mapsto \mathamsbb{R}^{p'}$, $\bpsi(\bx)$ is a signal 
transform-coefficient vector with most elements having negligible 
magnitudes.  The idea behind compressed sensing \cite{Candes2006} is to 
\emph{sense} the significant components of $\bpsi(\bx)$ using a small 
number of measurements. Define the noiseless measurement vector 
$\bm{\phi}(\bx)$, where $\bm{\phi}(\cdot): \mathamsbb{R}^p \mapsto 
\mathamsbb{R}^N$ and $N \leq p$.
Most effort in compressed sensing has focused on the linear sparsifying 
transform and noiseless measurement models with
\begin{subequations}
\begin{IEEEeqnarray}{c"c}
 \label{eq:Psibx}
  \bpsi(\bx)=\Psi^T\bx
\\
 \label{eq:measurelin}
  \bm{\phi}(\bx)=\Phi \bx
 \end{IEEEeqnarray}
 \end{subequations}
where $\Psi \in \mathamsbb{R}^{p \times p'}$  and $\Phi \in 
\mathamsbb{R}^{N \times p}$ are known  \emph{sparsifying dictionary}
and \emph{sensing} matrices.  Here, we consider signals $\bx$ that belong 
to a closed convex set $C$ in addition to their sparsity in the transform 
domain. The nonnegative signal scenario with
\begin{IEEEeqnarray}{c}
  \label{eq:nonneg}
  C = \mathamsbb{R}_+^p
\end{IEEEeqnarray}
is of significant practical interest and applicable to X-ray \gls{CT}, 
\gls{SPECT}, \gls{PET}, and \gls{MRI}, where the pixel values correspond to 
inherently nonnegative
density or concentration maps \cite{PrinceLinks2015}.
\citeauthor{Harmany2012} consider such a nonnegative sparse signal model 
and develop in \cite{Harmany2012} and \cite{Harmany2010Gradient} a 
convex-relaxation \gls{SPIRAL} and a linearly constrained gradient 
projection method for Poisson and Gaussian linear measurements, 
respectively.
In addition to signal nonnegativity, other convex-set constraints have been 
considered in the literature, such as prescribed value in the Fourier 
domain; box, geometric, and total-energy constraints; and intersections of 
these sets \cite{YoulaWebb1982,SezanStark1983}.

\begin{subequations}
  \label{eq:optgoal}
  We adopt the analysis regularization framework and minimize
  \begin{IEEEeqnarray}{c}
    \label{eq:f}
    f(\bx)=\cL(\bx) +u r(\bx)
  \end{IEEEeqnarray}
  with respect to the signal $\bx$, where $\cL(\bx)$ is a convex 
  differentiable \emph{data-fidelity} (\gls{NLL}) term,  $u > 0$ is a 
  scalar tuning constant that quantifies the weight of the convex 
  \emph{regularization term} $r(\bx)$ that
  imposes signal sparsity and the convex-set constraint:
  \begin{IEEEeqnarray}{c}
    \label{eq:r}
    r(\bx)=  \norm{\bpsi(\bx)}_1 + \mathamsbb{I}_C(\bx)
  \end{IEEEeqnarray}
\end{subequations}
 and
%
$\mathamsbb{I}_C(\ba) \df \ccases{ 0, & \ba \in C\\
    +\infty, & \text{otherwise}
  }$
is the indicator function. Common choices for the signal sparsifying 
transform are the linear map in \eqref{eq:Psibx}, isotropic gradient map
\begin{IEEEeqnarray}{c}
  \label{eq:gradientMap}
  \SBR{\bpsi(\bx)}_{i=1}^{p'} \df 
  \sqrt{\sum_{j\in\mathcal{N}_i}(x_i-x_j)^2}
\end{IEEEeqnarray}
and their combinations;  here, $\mathcal{N}_i$ is the index set of 
neighbors of $x_i$ in an appropriate (e.g., 2D) arrangement.  Summing 
\eqref{eq:gradientMap} over $i$ leads to the isotropic \gls{TV} penalty 
\cite{gdasil15,Harmany2012,Beck2009TV}; in the 2D case, anisotropic 
\gls{TV} penalty is slightly different and easy to accommodate as well.  
Assume
\begin{IEEEeqnarray}{c}
  \label{eq:Ccond}
  C \subseteq \closure\PARENSs{\dom \cL(\bx)}
\end{IEEEeqnarray}
which ensures that $\cL(\bx)$ is computable for all $\bx \in \interior C$ 
and closure ensures that points in $\dom\cL$ but close to its open 
boundary, if there is any, will not be excluded upon projecting onto the 
closed set $C$.   If
$C \setminus \dom \cL = C \cap \SBRbig{ \closure\PARENSs{\dom \cL(\bx)} 
\setminus \dom \cL  }$
is not empty, then $\cL(\bx)$ is not computable in it,  which needs special 
attention; see Section~\ref{sec:reconalg}.



  
 Define the proximal operator for a function $r(\bx)$ scaled by $\lambda$: 
\begin{IEEEeqnarray}{c}
  \label{eq:prox}
  \proxop_{\lambda r} \ba = \arg \min_{\bx}
  \tfrac{1}{2}\norm{\bx-\ba}_2^2 + \lambda r(\bx).
\end{IEEEeqnarray}
References \cite{DupeFadiliStarck2012,Raguet2013GFB,Condat2013Primal} view 
\eqref{eq:f} as a sum of three terms, $\cL(\bx)$, $\norm{\bpsi(\bx)}_1$, 
and $\mathamsbb{I}_C(\bx)$, and minimize it by splitting schemes, such as 
forward-backward, Douglas-Rachford, and primal-dual.  A potential benefit 
of splitting schemes is that they apply proximal operations on individual 
summands rather than on their combination, which is useful if all 
individual proximal operators are easy to compute.
However, \cite{DupeFadiliStarck2012} requires the proximal operator of 
$\cL(\bx)$, which is difficult in general and needs an inner iteration.  Both 
\cite{DupeFadiliStarck2012} and \gls{GFB} splitting \cite{Raguet2013GFB} 
require inner iterations for solving
$\proxs{\lambda \norm{\Psi^T \cdot }_1 }{\ba}$
(see \eqref{eq:Psibx} and \eqref{eq:prox})
in the general case where the sparsifying matrix $\Psi$ is not orthogonal.  
The elegant \gls{PDS} method in \cite{Condat2013Primal,Vu2013} does not 
require inner iterations.  The convergence rate of both \gls{GFB} and \gls{PDS} 
methods can be upper-bounded by $C/k$  where $k$ is the number of iterations 
and the constant $C$ is determined by values of the tuning proximal and 
relaxation constants \cite{Liang2014Conv, Davis2015ConvPDS}.



In this paper, we develop a \gls{PNPG} method whose momentum acceleration 
accommodates (increasing) adaptive step size selection (see also 
\cite{gdasil15,gdasil14,NesterovTechReport}) and convex-set constraint on 
the signal $\bx$.  \gls{PNPG} needs an inner iteration to compute the 
proximal operator with respect to $r(\bx)$, which implies inexact proximal 
operation.  We account for this inexactness and establish 
${\mathcal{O}(k^{-2})}$ convergence rate of the \gls{PNPG} method \emph{as 
well as} convergence of its iterates; the obtained convergence conditions 
motivate our selection of convergence criteria for proximal-mapping 
iterations.  We modify the original Nesterov's acceleration 
\cite{Nesterov1983,Beck2009FISTA} so that we can establish these 
convergence results when the step size is adaptive and
adjusts to the local curvature of the \gls{NLL}.
Thanks to the step-size adaptation, \gls{PNPG} \emph{does not} require
Lipschitz continuity of the gradient of the \gls{NLL} and applies to the 
Poisson compressed-sensing scenario described in 
Section~\ref{sec:PoissonGLM}.  Our integration of the adaptive step size 
and convex-set constraint extends the application of the Nesterov-type 
acceleration to more general measurement models than those used previously.  
Furthermore, a convex-set constraint can bring significant improvement to 
signal reconstructions compared with imposing signal sparsity only, as 
illustrated in Section~\ref{sec:linear1dex}.  See 
Section~\ref{sec:Okminustwoaccelerationapproaches} for discussion of other 
${\mathcal{O}(k^{-2})}$ acceleration approaches: Auslender-Teboulle (AT) 
\cite{Auslender2006AT,Becker2011TFOCS} 
and \citeauthor{BonettiniPortaRuggiero2015} 
\cite{BonettiniPortaRuggiero2015}.
Proximal Quasi-Newton type methods with \emph{problem-specific} diagonal 
Hessian approximations have been considered in 
\cite{BonettiniPortaRuggiero2015,BonettiniLorisPortaPrato2015}; 
\cite{BonettiniLorisPortaPrato2015} applies step-size adaptation and 
accounts for inaccurate proximal operator, but \emph{does not} employ 
acceleration or provide fast convergence-rate guarantees.

\gls{PNPG} code is easy to maintain: for example, the proximal-mapping 
computation can be easily replaced as a module by the latest 
state-of-the-art solver.  Furthermore, \gls{PNPG} requires minimal 
\emph{application-independent tuning};  indeed, we use the same set of 
tuning parameters in two different application examples.  This is in 
contrast with the existing splitting methods, which   require 
problem-dependent (\gls{NLL}-dependent) design and tuning.




We introduce the notation: $\bm{0}$, $\bm{1}$, $I$, denote the vectors of 
zeros and ones and identity matrix, respectively; ``$\succeq$'' is the 
elementwise version of ``$\geq$''.   For a vector $\bm{a} = 
\PARENSs{a_i}_{i=1}^N \in \mathamsbb{R}^{N \times 1}$, define the 
projection and soft-thresholding operators:
\begin{subequations}
  \begin{IEEEeqnarray}{rCl}
    \proj{C}{\ba} &=& \arg \min_{\bx \in C} \norm{\bx-\ba}_2^2\\
    \SBR{\softthr{\lambda}{\ba}}_i&=&\sgn(a_i)\maxp{ \abs{a_i} -\lambda,0 }
  \end{IEEEeqnarray}
and the elementwise logarithm and exponential functions $\SBRs{\ln_{\circ} 
\ba }_i = \ln a_i$ and $\SBRs{\exp_\circ \ba }_i= \exp a_i$.
The projection onto $\mathamsbb{R}_+^N$ and the proximal operator 
\eqref{eq:prox}
 for the $\ell_1$-norm $\|\bx\|_1$ can be computed in closed form:
\begin{IEEEeqnarray}{c"c}
  \label{eq:nonnegprojsoftthresh}
  \SBRbig{\proj{\mathamsbb{R}_+^N}{\ba}}_i =  \max(a_i,0), &
\proxop_{\lambda \norm{ \cdot }_1 } \ba  = \softthr{\lambda}{\ba}.
\end{IEEEeqnarray}
\end{subequations}

Define the $\varepsilon$-subgradient \cite[Sec.~23]{Rockafellar1970}:
\begin{IEEEeqnarray}{rCl}
  \label{eq:epsSubGrad}
  \partial_{\varepsilon}r(\bx) &\df& \CBRbig{ \bm{g}\in\mathamsbb{R}^p \mid  
  r(\bm{z})\geq r(\bx)+(\bz-\bx)^T\bm{g}-\varepsilon, \forall 
  \bz\in\mathamsbb{R}^p }
  \notag\\
\end{IEEEeqnarray}
and an \emph{inexact proximal operator}:
\begin{definition}
  \label{eq:ePrecision}
  We say that $\bx$ is an approximation of $\proxp{ur}{\ba}$ with 
  $\varepsilon$-precision \cite{Villa2013Inexact}, denoted
  \begin{subequations}
    \begin{IEEEeqnarray}{c}
      \bx\approxeq_\varepsilon \proxs{u r}{\ba}
      \label{eq:proximallambdaE}
    \end{IEEEeqnarray}
    if
    \begin{IEEEeqnarray}{c}
      \label{eq:esubr}
      \frac{\ba-\bx}{u} \in \partial_{\frac{\varepsilon^2}{2u}}r(\bx).
    \end{IEEEeqnarray}
  \end{subequations}
\end{definition}

Note that \eqref{eq:proximallambdaE} implies
\begin{IEEEeqnarray}{c}
  \label{eq:proximallambdaE2}
  \norm{\bx-\proxs{u r}{\ba}}_2^2\leq\varepsilon^2.
\end{IEEEeqnarray}

We introduce representative \gls{NLL} functions in 
Section~\ref{sec:measurementmodels}, describe the proposed \gls{PNPG} 
reconstruction algorithm in Section~\ref{sec:reconalg}, establish its 
convergence properties (Section~\ref{sec:convergence_analysis}), present 
numerical examples (Section~\ref{sec:NumEx}), and make concluding remarks 
(Section~\ref{sec:conclusion}).

\section{Probabilistic Measurement Models}
\label{sec:measurementmodels}

For numerical stability, we normalize the likelihood function so that the 
corresponding \gls{NLL} $\cL(\bx)$ is lower-bounded by zero.  For 
\glspl{NLL} that correspond to discrete \glspl{GLM},
 this normalization corresponds to the generalized Kullback-Leibler 
 divergence form of the \gls{NLL} and is also  closely related to the 
 Bregman divergence  \cite{BanerjeeDhillon2005}.

\subsection{Poisson Generalized Linear Model}
\label{sec:PoissonGLM}

\Glspl{GLM} with Poisson observations are often adopted in astronomic, 
optical, hyperspectral, and tomographic imaging 
\cite{Willett2014,PrinceLinks2015,StarckMurtagh2006,Snyder1993}  and used 
to model event counts, e.g., numbers of particles hitting a detector.  
Assume that the measurements $\by = \PARENSs{y_n}_{n=1}^N \in 
\mathamsbb{N}_0^N$ are independent Poisson-distributed\footnote{
  Here, we use the extended Poisson \gls{pmf}
  $\Poisson( y \,|\, \mu ) = \frac{\mu^y}{y!} \E^{-\mu}$ for all $\mu\geq0$ 
  by defining $0^0=1$ to accommodate the identity-link model.
}
with means $\SBRs{\bphi(\bx)}_n$.

Upon ignoring constant terms and normalization, we obtain the generalized 
Kullback-Leibler divergence form \cite{ZanniBertero2014} of the \gls{NLL}
\begin{subequations}
\begin{IEEEeqnarray}{c}
  \cL(\bx)=\bm{1}^T \SBRs{\bm{\phi}(\bx )-\by} + \sum_{n,y_n\neq0}y_n
  \ln \frac{y_n}{\SBRs{\bphi(\bx)}_n}.
  \label{eq:poissonl}
\end{IEEEeqnarray}
The \gls{NLL} $\cL(\bx): \mathamsbb{R}^p \mapsto \mathamsbb{R}_+$ is a convex 
function of the signal $\bx$.
Here, the relationship between the linear predictor $\Phi\bx$ and the 
expected value $\bphi(\bx)$ of the measurements $\by$ is summarized by the 
link function $\bg(\cdot): \mathamsbb{R}^N \mapsto \mathamsbb{R}^N$ 
\cite{McCullagh1989}:
\begin{IEEEeqnarray}{c}
  \Exp(\by) = \bphi(\bx)=\bg^{-1}(\Phi\bx).
\end{IEEEeqnarray}
\end{subequations}
Note that $\closure\PARENSs{\dom\cL}
=\CBRs{\bx\in\mathamsbb{R}^p \mid \bphi(\bx)\succeq\bm{0}}$.


Two typical link functions in the Poisson \gls{GLM} are log and identity, 
described in the following:

\subsubsection{Identity link}
\label{sec:idlink}
The identity link function with
\begin{IEEEeqnarray}{c"c}
  \label{eq:identityLink}
  \bg(\bm{\mu}) = \bm{\mu}-\bb, &
  \, \bphi(\bx) = \Phi\bx + \bb 
\end{IEEEeqnarray}
is used for modeling the photon count in optical imaging \cite{Snyder1993} 
and radiation activity in emission tomography 
\cite[Ch.~9.2]{PrinceLinks2015}, as well as for astronomical image 
deconvolution  \cite[Sec.~3.5.4]{StarckMurtagh2006}.  Here, $\Phi= \in 
\mathamsbb{R}_+^{N\times p}$ and $\bb \in \mathamsbb{R}_+^{N\times 1}$ are 
the known sensing matrix and intercept term, respectively; the intercept 
$\bb$ models background radiation and scattering determined, e.g., by 
calibration before the measurements $\by$ have been collected.  
The nonnegative set $C$ in \eqref{eq:nonneg} satisfies \eqref{eq:Ccond}, 
where we have used the fact that the elements of $\Phi$ are nonnegative.  
If $\bb$ has zero components, $C \setminus \dom \cL$ is \emph{not empty}  
and the \gls{NLL} does not have a Lipschitz-continuous gradient.

Setting $\bb=\bm{0}$ leads to the identity link without intercept used, 
e.g., in \cite{Snyder1993,StarckMurtagh2006,Harmany2012}.

\subsubsection{Log link}
\label{sec:Poissonloglink}

The log-link function
\begin{IEEEeqnarray}{c"c}
  \bg(\bm{\mu})
  =  -{\ln_\circ \PARENSs{ {\bm{\mu} }/{\cI_0 }}}, & \bphi(\bx) = 
  \cI_0\exp_\circ(-\Phi\bx)
  \label{eq:loglinear}
\end{IEEEeqnarray}
has been used to account for the exponential attenuation of particles 
(e.g., in tomographic imaging), where $\cI_0$ is the incident energy before 
attenuation.  The intercept term $\cI_0$ is often assumed known \cite[Sec.  
8.10]{Lange2013}.  The Poisson \gls{GLM} with log link function is referred 
to as the \emph{log-linear model} in \cite[Ch.~6]{McCullagh1989}, which 
treats  known and unknown $\cI_0$ as the same model.  


\textbf{Log link with unknown intercept.} For unknown $\cI_0$, 
\eqref{eq:poissonl} does not hold because the underlying \gls{NLL} is a 
function of \emph{both} $\bx$ and $\cI_0$.  Substituting 
\eqref{eq:loglinear} into the \gls{NLL} function, concentrating it with 
respect to $\cI_0$, and ignoring constant terms yields the following convex 
concentrated (profile) \gls{NLL}:
\begin{IEEEeqnarray}{c}
  \cL_\tc(\bx) = \bm{1}^T\by \ln\bigl[ \bm{1}^T\exp_\circ(-\Phi\bx) \bigr]
  + \by^T\Phi\bx
  \label{eq:concentratednllpoisson}
\end{IEEEeqnarray}
see 
\cite[App.~\ref{report-app:derconcentratednllpoisson}]{NesterovTechReport}, 
where we also derive the Hessian of \eqref{eq:concentratednllpoisson}.  
Note that $\dom\cL_\tc(\bx)=\mathamsbb{R}^p$; hence, any closed convex $C$ 
satisfies \eqref{eq:Ccond}.

\subsection{Linear Model with Gaussian Noise}
\label{sec:linGauss}

Linear measurement model
\eqref{eq:measurelin} with zero-mean \gls{AWGN}
leads to the following scaled
\gls{NLL}:
\begin{IEEEeqnarray}{c}
  \cL(\bx) = \frac{1}{2} \|\by-\Phi\bx\|_2^2
  \label{eq:gaussianL}
\end{IEEEeqnarray}
where $\by \in \mathamsbb{R}^N$ is the  measurement vector and constant terms 
(not functions of $\bx$) have been ignored.
This \gls{NLL} belongs to the Gaussian \gls{GLM} with identity link without 
intercept: $\bg(\bmu)=\bmu$.
Here, $\dom\cL(\bx)=\mathamsbb{R}^p$, any closed convex $C$ satisfies 
\eqref{eq:Ccond}, and the set $C \setminus \dom \cL$ is empty.

Minimization of the objective function \eqref{eq:f} with penalty 
\eqref{eq:r} and Gaussian \gls{NLL} \eqref{eq:gaussianL} can be thought of 
as an \emph{analysis \gls{BPDN} problem with a convex signal constraint}; 
see also \cite{NesterovTechReport,gdasil14} which use the nonnegative $C$ 
in \eqref{eq:nonneg}. A synthesis \gls{BPDN} problem with a convex signal 
constraint was considered in \cite{YamagishiYamada2009}.

\section{Reconstruction Algorithm}
\label{sec:reconalg}

We propose a \gls{PNPG} approach for minimizing \eqref{eq:f} that combines 
convex-set projection with Nesterov acceleration 
\cite{Nesterov1983,Beck2009FISTA} and applies
adaptive step size to adapt to the local curvature of the \gls{NLL} and
restart to ensure monotonicity of the resulting iteration. The pseudo code 
in Algorithm~\ref{alg:nesterov} summarizes our \gls{PNPG} method.


\begin{algorithm}[t]
  \DontPrintSemicolon
  \KwIn{$\bx^{(0)}$, $u$, $\gamma$, $b$, $\mathbb{n}$, $\mathbb{m}$, $\xi$, 
$\eta$, and threshold $\epsilon$}
  \KwOut{$\arg\min_\bx f(\bx)$}
  Initialization: $\step{\bx}{-1}\gets\bm{0}$, $i\gets 0$, $\kappa \gets 
  0$, $\beta^{(1)}$ by the BB method\;

  \Repeat{
    convergence declared or maximum number of iterations exceeded
  }{
    $i \gets i+1$ and $\kappa \gets \kappa+1$\;
    \While(\tcp*[f]{\small{backtracking search}}){true}{
      evaluate \eqref{eq:B} to \eqref{eq:nesterov}\;
      \If(\tcp*[f]{\small{domain restart}}){$\wbx^{(i)}\notin\dom\cL$}{
        $\theta^{(i-1)} \gets 1$ and continue\;
      }
      solve the proximal-mapping step \eqref{eq:proxgradstep}\;
      \eIf{majorization condition \eqref{eq:majorCond} holds}{
        break\;
      }{
        \If(\tcp*[f]{\small{increase $\mathbb{n}$}})
        {$\beta^{(i)}>\beta^{(i-1)}$}{
          $\mathbb{n} \gets \mathbb{n}+\mathbb{m}$
        }
        $\beta^{(i)} \gets \xi\beta^{(i)}$ and $ \kappa \gets 0$\;
      }
    }
    \If(\tcp*[f]{\small{restart}}){$i>1$ and $f\PARENSbig{\bx^{(i)}} > 
    f\PARENSbig{\bx^{(i-1)}}$}{
        $\theta^{(i-1)} \gets 1$, $i\gets i-1$, and continue
    }
    \If{convergence condition \eqref{eq:convcondouteriteration} holds with 
    threshold $\epsilon$
    }{
        declare convergence
    }
    \eIf(\tcp*[f]{\small{adapt step size}}){$\kappa \geq \mathbb{n}$}{
      $\kappa \gets 0$ and $\beta^{(i+1)} \gets {\beta^{(i)}}/{\xi}$
    }{
      $\beta^{(i+1)} \gets \beta^{(i)}$
    }
  }
  \caption{PNPG iteration}
  \label{alg:nesterov}
\end{algorithm}

Define the quadratic approximation of the \gls{NLL} $\cL(\bx)$:
\begin{IEEEeqnarray}{c}
  \label{eq:Q}
  Q_{\beta}\PARENSs{\bx \mid \wbx} = \mathcal{L}(\wbx)
  + (\bx-\wbx)^T\nabla\mathcal{L}(\wbx)
  + \frac{1}{2\beta}\norm{\bx-\wbx}_2^2
    \IEEEeqnarraynumspace
\end{IEEEeqnarray}
with $\beta$ chosen so that \eqref{eq:Q} majorizes $\cL(\bx)$ in the 
neighborhood of $\bx=\wbx$.
Iteration~$i$ of the \gls{PNPG} method proceeds as follows:
\begin{subequations}
  \label{eq:PNPG}
  \begin{IEEEeqnarray}{rCl}
    \label{eq:B}
    B^{(i)}&=& {\beta^{(i-1)}}/{\beta^{(i)}}\\
    \label{eq:thetaupdateMod}
    \theta^{(i)} &=& \ccases{
      1, & i\leq1\\
      \frac{1}{\gamma}+\sqrt{{b}+B^{(i)}\PARENSs{\theta^{(i-1)}}^2}, & i>1
    }
    \\
    \label{eq:Theta}
    \Theta^{(i)}&=&\PARENSbig{\theta^{(i-1)}-1}/{\theta^{(i)}}
    \\
    \label{eq:nesterov}
    \wbx^{(i)} &=& P_C\PARENSBig{
      \bx^{(i-1)} + \Theta^{(i)}\bigl( \bx^{(i-1)} - \bx^{(i-2)} \bigr)
    }
    \IEEEeqnarraynumspace
    \\
    \label{eq:proxgradstep}
    \bx^{(i)} &=& \proxpbig{\beta^{(i)} u r} { \wbx^{(i)} -
    \beta^{(i)} \nabla \cL( \wbx^{(i)} )}
  \end{IEEEeqnarray}
\end{subequations}
where
 $\beta^{(i)}>0$ is an \emph{adaptive step size} chosen to satisfy the 
 \emph{majorization condition}
\begin{IEEEeqnarray}{rCl}
  \mathcal{L}\PARENSs{\bx^{(i)}} &\leq&
  Q_{\beta^{(i)}}\PARENSs{ \bx^{(i)} \mid {\wbx}^{(i)} }
  \label{eq:majorCond}
\end{IEEEeqnarray}
using a simple adaptation scheme that aims at keeping $\step{\beta}{i}$ as 
large as possible, described in Section~\ref{sec:stepsize}; see also 
Algorithm~\ref{alg:nesterov}.
Here,
\begin{IEEEeqnarray}{c"c}
  \label{eq:gammab}
  \gamma\geq2, & b \in \SBRs{0,1/4}
\end{IEEEeqnarray}
in \eqref{eq:thetaupdateMod} are \emph{momentum tuning constants} and 
$\gamma$ controls the rate of increase of the extrapolation term 
$\theta^{(i)}$.  We will denote $\theta^{(i)}$ as $\theta_{\gamma,b}^{(i)}$ 
when we wish to emphasize its dependence on $\gamma$ and $b$. 

\begin{figure}
  \centering
  {\includegraphics{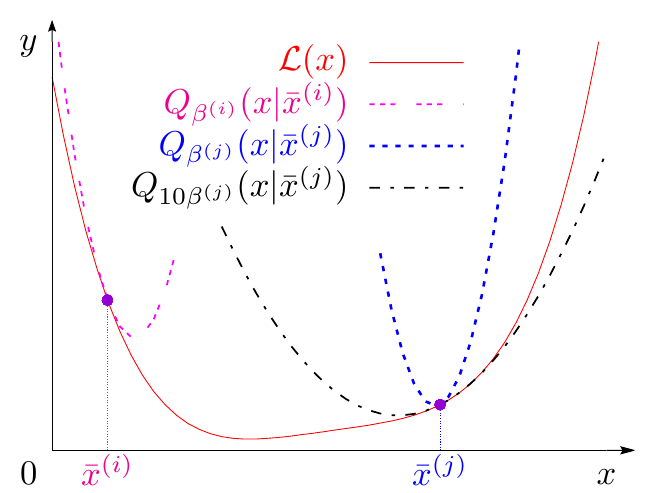}}
  \caption{Local and global majorizing functions.}
  \label{fig:localLip}
\end{figure}

Note that $Q_{\beta}\PARENSs{\bx \mid \wbx}$  is a \emph{local} majorizer 
of $\cL(\bx)$ in the neighborhood of $\wbx$: unlike most existing work, we 
require \eqref{eq:majorCond} to hold only for $\bx^{(i)}$, i.e., $ 
\mathcal{L}\PARENSs{\bx} \nleq Q_{\beta^{(i)}}\PARENSs{ \bx \mid 
  {\wbx}^{(i)} }$ in general, which violates the conventional textbook 
  definitions of majorizing functions that require \emph{global}, rather 
  than local, majorization; see also Fig.~\ref{fig:localLip} and 
  \cite[Sec.~5.8 and Fig.~5.10]{HastieTibshiraniWainwright2015}.
By relaxing the global majorization requirement, we aim to obtain a local 
majorizer such as $Q_{10\beta^{(j)}}$, which approximates the \gls{NLL} 
function \emph{better} in the neighborhood of $\step{\bar{\bx}}{j}$ than 
the global majorizer and also provides the same monotonicity guarantees as 
the global majorizer \emph{as long as} the majorization condition holds at 
the new iterate to which we move next.


For $\bpsi(\cdot)$ in \eqref{eq:Psibx} and \eqref{eq:gradientMap}, we 
compute the proximal mapping \eqref{eq:proxgradstep} using the \gls{ADMM} 
in Section~\ref{sec:proximalmapping} and
 \gls{TV}-based denoising method in \cite{Beck2009TV}, respectively.  
 Because of its iterative nature,  \eqref{eq:proxgradstep} is 
 \emph{inexact}; this inexactness can be modeled as
 \begin{IEEEeqnarray}{c}
    \label{eq:proxgradstepE}
    \bx^{(i)} \approxeq_{\varepsilon^{(i)}}
    \proxpbig{\beta^{(i)} u r} { \wbx^{(i)} -
    \beta^{(i)} \nabla \cL( \wbx^{(i)} )}
  \end{IEEEeqnarray}
  where $\varepsilon^{(i)}$ quantifies the precision of the \gls{PG} step 
  in Iteration~$i$.

\begin{figure*}
  \centering
  \def\width{0.43}
  \begin{subfigure}[c]{\width\textwidth}
    \centering
    \makebox[0pt][c]{{\includegraphics{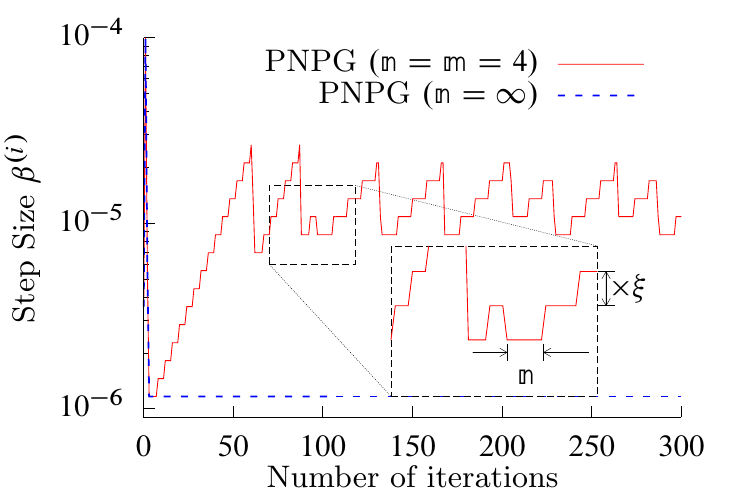}}}
    \caption{Poisson model with identity link.}
    \label{fig:stepSizePoi}
  \end{subfigure}
  \begin{subfigure}[c]{\width\textwidth}
    \centering
    \makebox[0pt][c]{{\includegraphics{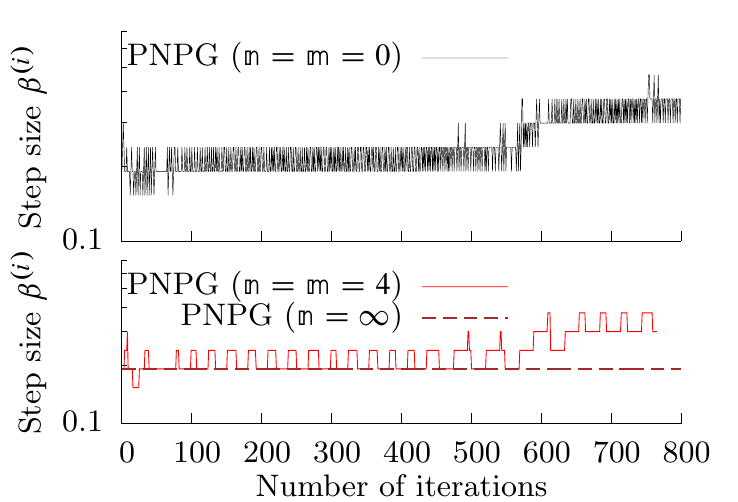}}}
    \caption{Gaussian linear model.}
    \label{fig:stepSizeLin}
  \end{subfigure}
   \caption{
    Step sizes $\beta^{(i)}$ as functions of the number of iterations for 
    Poisson and Gaussian linear models.
  }
  \label{fig:stepSize}
\end{figure*}

The acceleration step \eqref{eq:nesterov}  extrapolates the two latest 
iteration points in the direction of their difference $\bx^{(i)} - 
\bx^{(i-1)}$, followed by the projection onto the convex set $C$.  
For nonnegative $C$ in \eqref{eq:nonneg}, this projection has closed form, 
see \eqref{eq:nonnegprojsoftthresh}.  If $C$ is an intersection of convex 
sets with simple individual projection operator for each, we can apply 
\gls{POCS} \cite{YoulaWebb1982}.

If we remove the signal-sparsity penalty term by setting 
$\bpsi(\cdot)=\bm{0}$, the proximal mapping in \eqref{eq:proxgradstep} 
reduces to the projection onto $C$ and the iteration \eqref{eq:PNPG} 
becomes a projected Nesterov's \emph{projected} gradient method: an 
approach to accelerate projected gradient methods that ensures that the 
gradient of the \gls{NLL} is computable at the extrapolated value 
$\wbx^{(i)}$.  When projection onto $C$ is simple, a suboptimal \emph{ad 
hoc} approach that avoids inner proximal-mapping iteration can be 
constructed as follows: replace the first summand in \eqref{eq:r} with a 
differentiable approximation and then append the resulting approximate term 
to the \gls{NLL} term, thus leaving $r(\bx)$ with the indicator function 
only, which leads to the projected Nesterov's projected gradient method.  A 
similar approach has been used in \cite{BonettiniPortaRuggiero2015} in its 
Poisson image deblurring numerical example.


If we remove the convex-set constraint by setting $C=\mathamsbb{R}^p$, 
iteration \eqref{eq:B}--\eqref{eq:proxgradstep} reduces to the Nesterov's 
proximal-gradient iteration with adaptive step size that imposes signal 
sparsity \emph{only} in the analysis form (termed \gls{NPGS}); see also 
Section~\ref{sec:linear1dex} for an illustrative comparison of \gls{NPGS} 
and \gls{PNPG}. 

We now extend \cite[Lemma~2.3]{Beck2009FISTA} to the inexact proximal 
operation:
\begin{lem}
  \label{thm:lll}
  Assume convex and differentiable \gls{NLL} $\cL(\bx)$ and convex $r(\bx)$ 
  and
  consider an inexact \gls{PG} step \eqref{eq:proxgradstepE}
  with step size $\beta^{(i)}$ that satisfies the majorization condition 
  \eqref{eq:majorCond}.  Then,
  \begin{IEEEeqnarray}{rCl}
    f(\bx) - f(\bx^{(i)})
    &\geq&
    \frac{1}{2\beta^{(i)}}
    \bigl[
      \norm{\bx^{(i)}-\bx}_2^2
      -\norm{\wbx^{(i)}-\bx}_2^2
      -\PARENSs{\varepsilon^{(i)}}^2
    \bigr]
    \notag\\
    \label{eq:lem23}
  \end{IEEEeqnarray}
  for all $i\geq1$ and any $\bx\in\mathamsbb{R}^p$.
\end{lem}
\begin{IEEEproof}
  See Appendix~\ref{app:deracceleration}.
\end{IEEEproof}

Lemma~\ref{thm:lll} is general and algorithm-independent because 
$\wbx^{(i)}$ can be any value in $\dom\cL$,
the regularization term $r(\bx)$ can be any convex function,
and we have used only the fact that step size $\step{\beta}{i}$ satisfies 
the majorization condition \eqref{eq:majorCond}, rather than specific 
details of the step-size selection.  We will use this result  to establish 
the monotonicity property in Remark~\ref{lemma:monotonic} and as the 
starting point for deriving and analyzing our accelerated \gls{PG} scheme. 

\subsection{Restart and Monotonicity}
\label{sec:restartmonotonicity}

If $f\PARENSs{\bx^{(i)}}>f\PARENSs{\bx^{(i-1)}}$ or $\wbx^{(i)}\in 
C\setminus\dom\cL$, set
\begin{IEEEeqnarray}{c"s}
  \label{eq:restart}
  \theta^{(i-1)}=1 & (restart)
\end{IEEEeqnarray}
and refer to this action as
\emph{function restart} \cite{ODonoghue2013}
or \emph{domain restart} respectively;  see Algorithm~\ref{alg:nesterov}.
The goal of function and domain restarts is to ensure that the \gls{PNPG} 
iteration is monotonic and $\wbx^{(i)}$ remains within $\dom f$ as long as 
the projected initial value is within $\dom f$: 
$f\PARENSbig{\proj{C}{\step{\bx}{0}}}<+\infty$.  

The majorization condition \eqref{eq:majorCond} ensures that the iterate 
$\step{\bx}{i}$ attains lower (or equal) objective function than the 
intermediate signal $\step{\wbar{\bx}}{i}$ (see Lemma~\ref{thm:lll} with 
$\step{\wbx}{i}$ in place of $\bx$)
\begin{IEEEeqnarray}{c}
  \label{eq:majorizationcondconsequence}
  f\PARENSs{\step{\bx}{i}} \leq
  f\PARENSs{\step{\wbar{\bx}}{i}}-\frac{1}{2\beta^{(i)}}
  \SBRbig{
    \norm{\step{\bx}{i}-\step{\wbar{\bx}}{i}}^2_2
    -\PARENSs{\varepsilon^{(i)}}^2
  }
  \IEEEeqnarraynumspace
\end{IEEEeqnarray}
provided that proximal-mapping approximation error term 
$\PARENSs{\varepsilon^{(i)}}^2$ is sufficiently small.
However, \eqref{eq:majorizationcondconsequence}
\emph{does not} guarantee monotonicity of the \gls{PNPG} iteration: we  
apply the function restart to restore the monotonicity and improve 
convergence of the \gls{PNPG} iteration; see Algorithm~\ref{alg:nesterov}.

Define the local variation of signal iterates
\begin{IEEEeqnarray}{c}
    \label{eq:deltai}
    \delta^{(i)} \df  \norm{\step{\bx}{i}-\step{\bx}{i-1}}_2^2.
  \end{IEEEeqnarray}

\begin{rem}[Monotonicity]
  \label{lemma:monotonic}
  The \gls{PNPG} iteration with restart and inexact \gls{PG} steps 
  \eqref{eq:proxgradstepE}  is non-increasing:
  \begin{subequations}
  \begin{IEEEeqnarray}{c}
  f\PARENSs{\bx^{(i)}} \leq f(\bx^{(i-1)})
  \label{eq:remmonotonicity}
\end{IEEEeqnarray}
for all $i$ if the inexact \gls{PG} steps  are
sufficiently accurate and satisfy
  \begin{IEEEeqnarray}{c}
    \label{eq:moncond}
    \varepsilon^{(i)}\leq \sqrt{\delta^{(i)}}.
  \end{IEEEeqnarray}
\end{subequations}
 \end{rem}
\begin{IEEEproof}
  If there is no restart in Iteration~$i$, the objective function has not 
  increased.  If there is a restart, $\theta^{(i-1)}=1$ and 
  \eqref{eq:nesterov} simplifies to
 $\wbx^{(i)} = \projpbig{C}{  \bx^{(i-1)} }  = \bx^{(i-1)}$, and 
 monotonicity follows upon substituting $\wbx^{(i)} = \bx^{(i-1)}$ into 
 \eqref{eq:majorizationcondconsequence}.\looseness=-1
\end{IEEEproof}
To establish the monotonicity in Remark~\ref{lemma:monotonic}, we only need 
the step size $\step{\beta}{i}$ to satisfy the majorization condition 
\eqref{eq:majorCond}. Small $\varepsilon^{(i)}$ can be achieved by 
selecting the inner-iteration convergence criteria appropriately; see 
Section~\ref{sec:convCrit}.

The condition \eqref{eq:moncond} motivates us to incorporate 
$\delta^{(i-1)}$ into the convergence criteria for proximal-step 
computation, see Section~\ref{sec:convCrit}.

\subsection{Adaptive Step Size}
\label{sec:stepsize}
Now, we present an adaptive scheme for selecting $\step{\beta}{i}$:
\begin{subequations}
  \label{eq:stepsizeadaptation}
  \begin{enumerate}[label=\roman*)]
    \item
      \begin{itemize}
        \item if there have been no step-size backtracking events or 
          increase attempts for $\mathbb{n}$ consecutive iterations 
          ($i-\mathbb{n}$ to $i-1$), start with a larger step size
          \begin{IEEEeqnarray}{c"s}
            \label{eq:increasedstepsize}
            \wbar{\beta}^{(i)} = {\beta^{(i-1)}} / {\xi}
            & (increase attempt)
          \end{IEEEeqnarray}
          where
          \begin{IEEEeqnarray}{c"s}
            \label{eq:xi}
          \xi \in (0,1)
          \end{IEEEeqnarray}
          is a \emph{step-size adaptation parameter};
        \item otherwise start with
          \begin{IEEEeqnarray}{c}
            \label{eq:oldstepsize}
            \wbar{\beta}^{(i)}=\beta^{(i-1)};
          \end{IEEEeqnarray}
      \end{itemize}
    \item \label{backtrack} (backtracking search) select
      \begin{IEEEeqnarray}{c}
        \label{eq:betai}
        \beta^{(i)} =\xi^{t_i}\wbar{\beta}^{(i)}
      \end{IEEEeqnarray}
      where $t_i\geq0$ is the smallest integer  such that \eqref{eq:betai} 
      satisfies the majorization condition \eqref{eq:majorCond}; 
      \emph{backtracking event} corresponds to $t_i>0$.
    \item if $\max\PARENSs{\beta^{(i)},\beta^{(i-1)}}<\wbar{\beta}^{(i)}$,  
      increase $\mathbb{n}$ by  a nonnegative integer $\mathbb{m}$:
      \begin{IEEEeqnarray}{rCl}
        \mathbb{n} \gets \mathbb{n}+\mathbb{m}.
      \end{IEEEeqnarray}
  \end{enumerate}
\end{subequations}
We select the initial step size $\wbar{\beta}^{(1)}$ using the \gls{BB} 
method \cite{BarzilaiBorwein1988}.

If there has been an attempt to change the step size in any of the previous 
$\mathbb{n}$ consecutive iterations, we start the backtracking search 
\ref{backtrack} with the step size from the latest completed iteration.  
Consequently, the step size will be approximately piecewise-constant as a 
function of the iteration index $i$; see Fig.~\ref{fig:stepSize}, which 
shows the evolutions of the adaptive step size $\step{\beta}{i}$ for 
measurements following the Poisson generalized linear and Gaussian linear 
models corresponding to Figs.~\ref{fig:cost_timePET} and 
\ref{fig:traceLinGaussNeg3}; see Sections~\ref{sec:poissonPET} and 
\ref{sec:linear1dex} for details of the two simulation scenarios.  Here, 
$\mathbb{n}$ controls the size of the neighborhood around $\step{\wbx}{i}$ 
over which $Q_{\beta^{(i)}}\PARENSbig{ \bx \mid {\wbx}^{(i)} }$ majorizes 
$\cL(\bx)$ and also the time spent backtracking: larger $\mathbb{n}$ yields 
a larger neighborhood and leads to less backtracking.  To reduce 
sensitivity to the choice of the tuning constant $\mathbb{n}$, we adapt it 
by increasing its value by $\mathbb{m}$ if there is a failed attempt to 
increase the step size in Iteration~$i$, i.e., 
$\wbar{\beta}^{(i)}>\beta^{(i-1)}$ and $\beta^{(i)}<\wbar{\beta}^{(i)}$.


The adaptive step-size strategy keeps $\beta^{(i)}$ as large as possible 
subject to \eqref{eq:majorCond}, which is important not only because the 
signal iterate may reach regions of $\cL(\bx)$ with different local 
Lipschitz constants, but also due to the varying curvature of $\cL(\bx)$ in 
different updating direction.  For example, a (backtracking-only) 
\gls{PG}-type algorithm with non-adaptive step size would fail or converge 
very slowly if the local Lipschitz constant of $\nabla\cL(\bx)$ decreases 
as the algorithm iterates because the step size will not be able to adjust and 
track this decrease; see also Section~\ref{sec:NumEx} where the benefits of 
step-size adaptation are demonstrated by numerical examples.

Setting $\mathbb{n}=+\infty$ corresponds to step-size backtracking only, 
which aims at finding $\step{\beta}{i}/\xi$ upper bounding the inverse of 
the (global) Lipschitz constant of $\nabla\cL(\bx)$.  A step-size 
adaptation scheme with $\mathbb{n}=\mathbb{m}=0$ initializes the step-size 
search aggressively, with an increase attempt \eqref{eq:increasedstepsize} 
in each iteration.


\subsection{ADMM Proximal Mapping for $\ell_1$-Norm Penalty with Linear 
Transform Coefficients}
\label{sec:proximalmapping}

We present an \emph{\gls{ADMM}} scheme for computing the proximal operator 
\eqref{eq:prox} with $\bpsi(\bx)$ in \eqref{eq:Psibx}:
\begin{subequations}
  \label{eq:ADMM}
  \begin{IEEEeqnarray}{rCl}
    \label{eq:abar}
    \bar{\ba}^{(i)}&=&
    \PARENSbig{I+\rho\Psi\Psi^T}^{-1}
    \SBRbig{
      \ba+\rho\Psi\PARENSs{ \step{\bs}{j-1}+\step{\bupsilon}{j-1} }
    }
    \\
    \label{eq:truncate}
    \step{\balpha}{j} &=&
    \arg\min_{\balpha\in C} 
    \normbig{\balpha-\bar{\ba}^{(i)}}_{I+\rho\Psi\Psi^T}^2
    \\
    \label{eq:sUpdate}
    \step{\bs}{j} &=& \mathcal{T}_{{\lambda}/{\rho}}
    \PARENSbig{\Psi^T\step{\balpha}{j}-\step{\bupsilon}{j-1}}\\
    \label{eq:dualUpdate}
    \step{\bupsilon}{j} &=&
    \step{\bupsilon}{j-1}+\step{\bs}{j}-\Psi^T\step{\balpha}{j}
  \end{IEEEeqnarray}
\end{subequations}
where $j$ is the \gls{ADMM} iteration index and $\rho>0$ is a
penalty parameter.  We obtain \eqref{eq:ADMM}  by decomposing 
\eqref{eq:prox} into the sum of 
$\frac{1}{2}\normlr{\balpha-\ba}_2^2+\mathamsbb{I}_C(\balpha)$ and $\lambda 
\norm{\bs}_1$ with equality constraints $\Psi^T \balpha=\bs$ 
\cite{Boyd2011ADMM}.
We initialize $\rho$ by 1 and adaptively adjust its value thereafter
using the scheme in \cite[Sec.~3.4.1]{Boyd2011ADMM}. We also set  
$\step{\bupsilon}{0}=\bm{0}_{p' \times 1}$.
The signal estimate $\balpha^{(j)}$ is returned upon convergence.


To apply the iteration  \eqref{eq:truncate}--\eqref{eq:dualUpdate} in 
large-scale problems, we need a computationally efficient solution to the 
linear system in \eqref{eq:truncate}; for example, we may exploit a special 
structure of $\Psi\Psi^T$ or use its accurate (block)-diagonal 
approximation.  In the absence of efficient or accurate approximate 
solutions, we can replace the full iterative solver of this linear system 
with its single step \cite[Sec.~4.4.2]{Parikh2013Proximal}.  In many 
applications, the rows of $\Psi$ are orthonormal:
\begin{IEEEeqnarray}{c}
  \label{eq:Psiorthogonalrows}
  \Psi\Psi^T=I
\end{IEEEeqnarray}
and \eqref{eq:truncate} simplifies greatly because 
$\PARENSs{I+\rho\Psi\Psi^T}^{-1}$ can be replaced with the scalar term 
$1/(1+\rho)$.
For comparison, \gls{SPIRAL}\footnote{\gls{SPIRAL} is a \gls{PG} method for 
solving \eqref{eq:prox} that employs \gls{BB} step size,  implements the 
$\ell_1$-norm by splitting $\Psi^T\bx$ to positive and negative components, 
and solves the resulting problem by a Lagrangian method; see 
Section~\ref{sec:NumEx} for examples of its performance.}
\cite{Harmany2012} requires orthogonal $\Psi$: 
$\Psi\Psi^T=\Psi^T\Psi=I$, which is more restrictive than 
\eqref{eq:Psiorthogonalrows}.

\subsubsection{Inner ADMM Iteration}
When computing the proximal operator in \eqref{eq:proxgradstep}, we start 
the inner iteration \eqref{eq:ADMM} for the $i$th outer iteration by 
$\step{\bs}{i,0}=\Psi^T\bx^{(i-1)}$.  Here and in the following section, we  
denote \eqref{eq:truncate} and \eqref{eq:sUpdate} using 
$\step{\balpha}{i,j}$ and $\step{\bs}{i,j}$ to emphasize their dependence 
on the outer iteration index~$i$.  Although \gls{ADMM} may converge slowly 
to a very accurate point, it can reach sufficient accuracy within tens of 
iterations \cite[Sec.~3.2.2]{Boyd2011ADMM}.  As an inner iteration, 
\gls{ADMM} \emph{does not} need to solve the \gls{PG} step with high 
accuracy.  Indeed, Theorem~\ref{thm:conv} in 
Section~\ref{sec:convergence_analysis} provides fast overall convergence 
guarantees for \gls{PNPG} iteration even when using inexact proximal 
operators.  We use this insight to select the inner-iteration convergence 
criterion \eqref{eq:admmCrit} that becomes gradually more stringent as the 
iteration proceeds; see Section~\ref{sec:convCrit}.

\subsection{Convergence Criteria}
\label{sec:convCrit}

The outer- and inner-iteration convergence criteria are
\begin{IEEEeqnarray}{c}
  \sqrt{\delta^{(i)}} \leq
  \epsilon {\norm{\step{\bx}{i}}}_2
  \label{eq:convcondouteriteration}
\end{IEEEeqnarray}
and
\begin{subequations}
  \label{eq:innerCrit}
  \begin{IEEEeqnarray}{u+rCl}
    \label{eq:TVcrit}
    TV: &
    \norm{ \step{\bx}{i,j}-\step{\bx}{i,j-1} }_2
    &\leq& \eta \sqrt{\delta^{(i-1)}}\\
    $\ell_1$: &
    \IEEEeqnarraymulticol{3}{l}{
      \max\PARENSbig{
        \norm{ \step{\bs}{i,j}-\Psi^T\step{\balpha}{i,j} }_2,
        \norm{ \step{\bs}{i,j}-\step{\bs}{i,j-1} }_2
      }
    } \notag \\
    &&\leq& \eta
    \norm{\Psi^T\PARENSs{\step{\bx}{i-1}-\step{\bx}{i-2}}}_2
    \IEEEeqnarraynumspace
    \label{eq:admmCrit}
  \end{IEEEeqnarray}
\end{subequations}
where $\epsilon>0$ is the convergence threshold, $\eta \in(0,1)$ is the 
inner-iteration convergence tuning constant, and $i$ and $j$ are the outer 
and inner iteration indices.  In practice, the convergence threshold on the 
right-hand side of \eqref{eq:admmCrit} can be relaxed to 
$\eta\norm{\Psi}_2\sqrt{\delta^{(i-1)}}$ when the spectral norm 
$\norm{\Psi}_2$ is known.


 The convergence tuning constant $\eta$ is chosen to trade off the accuracy 
 and speed of the inner iterations and provide sufficiently accurate 
 solutions to \eqref{eq:proxgradstep}.  Here, $\bs^{(i,j)}$ in 
 \eqref{eq:admmCrit} is the dual variable in our \gls{ADMM} iteration and 
 the criterion \eqref{eq:admmCrit} applies to the larger of the primal and 
 dual residuals $\| \step{\bs}{i,j}-\Psi^T\step{\balpha}{i,j} \|_2$ and $\| 
 \step{\bs}{i,j}-\step{\bs}{i,j-1} \|_2$
\cite[Sec.~3.3]{Boyd2011ADMM}.

The monotonicity of the \gls{PNPG} iteration with inexact \gls{PG} steps in 
Remark~\ref{lemma:monotonic} is guaranteed if the \gls{PG} steps are 
sufficiently accurate, i.e., \eqref{eq:moncond} holds.  We now describe our 
adjustment of the inner-iteration convergence tuning constant $\eta$ in 
\eqref{eq:admmCrit} that ensures monotonicity of this iteration.
%
According to Remark~\ref{lemma:monotonic}, 
$f\PARENSs{\bx^{(i)}}>f\PARENSs{\bx^{(i-1)}}$ implies 
$\varepsilon^{(i)}>\sqrt{\delta^{(i)}}$, i.e., $\bx^{(i)}$ is not 
sufficiently accurate.  So, we decrease $\eta$ 10 times and re-evaluate 
\eqref{eq:B}--\eqref{eq:proxgradstep} when the objective function increases 
in the step following a function restart, thus necessitating a consecutive 
function restart.  Hence, $\eta$ is decreased only in the rare event of 
multiple consecutive restarts.  
This adjustment reduces the dependence of the algorithm on the initial 
value of $\eta$.



\section{Convergence Analysis}
\label{sec:convergence_analysis}


We now bound the convergence rate of the \gls{PNPG} method without restart.

\begin{thm}[Convergence of the Objective Function]
  \label{thm:conv}
  Assume that the \gls{NLL} $\cL(\bx)$ is convex and differentiable, 
  $r(\bx)$ is convex, the closed convex set $C$ satisfies
  \begin{IEEEeqnarray}{rCl}
    \label{eq:CindomL}
    C\subseteq \dom \cL
  \end{IEEEeqnarray}
  (implying no need for domain restart) and \eqref{eq:gammab} holds.  
  Consider the \gls{PNPG} iteration without 
  restart where \eqref{eq:proxgradstep} in Iteration~$i$ is replaced with 
  the inexact \gls{PG} step in \eqref{eq:proxgradstepE}.  The convergence 
  rate
  of the \gls{PNPG} iteration is bounded as follows: for $k \geq 1$,
  \begin{subequations}
    \label{eq:objRate}
    \begin{IEEEeqnarray}{rCl}
      \label{eq:DueTo0Theta}
      \Delta^{(k)} &\leq&   \frac{
        \norm{\step{\bx}{0}-\bx^\star}^2_2
        +\mathcal{E}^{(k)}
      }{2\beta^{(k)}\PARENSs{\step{\theta}{k}}^2}
      \\
      &\leq&
      \label{eq:upperboundonDeltawithbetaonly}
      \gamma^2
      \frac{
        \norm{\step{\bx}{0}-\bx^\star}^2_2
        +\mathcal{E}^{(k)}
      }{2\PARENSbig{\sqrt{\beta^{(1)}}+\sum_{i=1}^k\sqrt{{\beta^{(i)}}}}^2}
    \end{IEEEeqnarray}
  \end{subequations}
where
  \begin{subequations}
    \begin{IEEEeqnarray}{rCl}
 \label{eq:xstar}
    \bx^\star &=& \arg \min_{\bx} f(\bx)\\
      \label{eq:Delta}
      \Delta^{(k)}&\df& f\PARENSs{\bx^{(k)}}-f\PARENSs{\bx^\star}
      \\
      \label{eq:E}
      \mathcal{E}^{(k)}
      &\df& \sum_{i=1}^k\PARENSbig{\theta^{(i)}\varepsilon^{(i)}}^2
    \end{IEEEeqnarray}
  \end{subequations}
  are the minimum point of $f(\bx)$, centered objective function and 
  cumulative error term that accounts for the inexact \gls{PG} steps 
  \eqref{eq:proxgradstepE}.
\end{thm}
\begin{IEEEproof}
  We outline the main steps of the proof; see 
  Appendix~\ref{app:deracceleration} for details.  The proof and derivation 
  of the projected Nesterov's acceleration step \eqref{eq:nesterov} are 
  inspired by but more general than \cite{Beck2009FISTA}: we start from 
  \eqref{eq:lem23} with $\bx$ replaced by $\bx = \bx^\star$ and $\bx = 
  \step{\bx}{i-1}$,
  \begin{subequations}
    \label{eq:twoineq}
    \begin{IEEEeqnarray}{rCl}
      \label{eq:stari}
      -\step{\Delta}{i} & \geq &
      \frac{
        \norm{\bx^{(i)}-\bx^\star}_2^2
        -\norm{\wbx^{(i)}-\bx^\star}_2^2
        -\PARENSs{\varepsilon^{(i)}}^2
      }
      {2\beta^{(i)}}
      \IEEEeqnarraynumspace
      \\
      \label{eq:im1i}
      \step{\Delta}{i-1} -\step{\Delta}{i}
      & \geq &
      \frac{
        \delta^{(i)}
        -\norm{\wbx^{(i)}-\bx^{(i-1)}}_2^2
        -\PARENSs{\varepsilon^{(i)}}^2
      }
      {2\beta^{(i)}}
    \end{IEEEeqnarray}
  \end{subequations}
  and design two coefficient sequences that multiply \eqref{eq:stari} and 
  \eqref{eq:im1i}, respectively, which ultimately leads to 
  \eqref{eq:B}--\eqref{eq:nesterov} and the convergence-rate guarantee in 
  \eqref{eq:DueTo0Theta}.  One set of boundary conditions on the 
  coefficient sequences leads to the projected momentum acceleration step 
  \eqref{eq:nesterov} and a second set of conditions leads to the following 
  inequality involving the momentum terms 
  $\step{\theta}{i-1},\step{\theta}{i}$ and step sizes 
  $\step{\beta}{i-1},\step{\beta}{i}$ for $i>1$:
\begin{subequations}
      \label{eq:thetaCond0}
    \begin{IEEEeqnarray}{c}
      \label{eq:thetaCond}
      \step{\beta}{i-1}{\PARENSs{\step{\theta}{i-1}}^2}
      \geq
      \step{\beta}{i}\SBRbig{\PARENSs{\step{\theta}{i}}^2-\step{\theta}{i}}
    \end{IEEEeqnarray}
    which implies
    \begin{IEEEeqnarray}{rCl}
      \label{eq:solvedTheta}
      \step{\theta}{i}
      &\leq&
      \tfrac{1}{2}+
      \sqrt{\tfrac{1}{4}+ \step{B}{i}
      \PARENSs{\step{\theta}{i-1}}^2}
    \end{IEEEeqnarray}
  \end{subequations}
  and allows us to construct the recursive update of $\step{\theta}{i}$ in 
  \eqref{eq:thetaupdateMod}, see Appendix~\ref{sec:goal0}.  Comparing 
  \eqref{eq:thetaupdateMod} with   \eqref{eq:solvedTheta}
  justifies the constraints in \eqref{eq:gammab}. We handle the projection 
  onto the closed convex set $C$ by using the nonexpansiveness of the 
  convex-set projection.
  Finally, \eqref{eq:upperboundonDeltawithbetaonly} follows from 
  \eqref{eq:DueTo0Theta} by using
  \begin{subequations}
    \begin{IEEEeqnarray}{rCl}
      \label{eq:grb1}
      \theta^{(k)}\sqrt{\beta^{(k)}}
      &\geq&
      \frac{1}{\gamma}\sqrt{\beta^{(k)}}+\theta^{(k-1)} \sqrt{\beta^{(k-1)}}
      \\
      \label{eq:thetaGrow}
      &\geq&
      \frac{1}{\gamma}\sum_{i=2}^k\sqrt{\beta^{(i)}}
      +\theta^{(1)} \sqrt{\beta^{(1)}}
    \end{IEEEeqnarray}
    for all $k>1$, where \eqref{eq:grb1} follows from the definitions of 
    $B^{(k)}$ and $\theta^{(k)}$ in  \eqref{eq:B} and 
\eqref{eq:thetaupdateMod},
and \eqref{eq:thetaGrow} follows by repeated application of the inequality 
\eqref{eq:grb1} with $k$ replaced by $k-1,k-2,\ldots,2$.  
\end{subequations}
\end{IEEEproof}

Theorem~\ref{thm:conv} shows that better initialization, smaller 
proximal-mapping approximation error, and larger step sizes 
$\PARENSs{\step{\beta}{i}}_{i=1}^k$ help lower the convergence-rate upper 
bounds in \eqref{eq:objRate}.  This result motivates our step-size 
adaptation with goal to maintain large $\PARENSs{\step{\beta}{i}}_{i=1}^k$, 
see Section~\ref{sec:stepsize}.

The assumptions of Theorem~\ref{thm:conv} are more general than those in 
Section~\ref{sec:introduction}; indeed the regularization term $r(\bx)$ can 
be any convex function.
To derive this theorem, we have used only the fact that step size 
$\step{\beta}{i}$ satisfies the majorization condition 
\eqref{eq:majorCond}, rather than specific details of step-size selection.  

To minimize the upper bound in \eqref{eq:DueTo0Theta}, we can select 
$\step{\theta}{i}$ to satisfy \eqref{eq:solvedTheta} with equality, which 
corresponds to $\theta_{2,1/4}^{(i)}$ in \eqref{eq:thetaupdateMod}, on the 
boundary of the feasible region in \eqref{eq:gammab}.
By \eqref{eq:grb1}, $\sqrt{\beta^{(k)}}\theta^{(k)}$ and the denominator of 
the bound in \eqref{eq:DueTo0Theta} are strictly increasing sequences.
The upper bound in \eqref{eq:upperboundonDeltawithbetaonly} is not a 
function of $b$ and is minimized with respect to $\gamma$ for $\gamma=2$  
given the fixed step sizes $\PARENSs{\step{\beta}{i}}_{i=0}^{+\infty}$;
it also decreases at the rate $\mathcal{O}(k^{-2})$:
\begin{cor}
  \label{cor:rateThetaMin}
  Under the assumptions of Theorem~\ref{thm:conv}, the convergence of 
  \gls{PNPG} iterates $\bx^{(k)}$ without restart is bounded as follows:
  \begin{subequations}
    \begin{IEEEeqnarray}{rCl}
      \label{eq:objRateBetaMin}
      \Delta^{(k)} \leq \gamma^2
      \frac{
        \norm{\step{\bx}{0}-\bx^\star}^2_2
        +\mathcal{E}^{(k)}
      }{2(k+1)^2\beta_\tmin}
    \end{IEEEeqnarray}
    for $k\geq1$, provided that
    \begin{IEEEeqnarray}{c}
      \label{eq:betaminpositive}
      \beta_\tmin \df \min_{k=1}^{+\infty} \step{\beta}{i} > 0.
    \end{IEEEeqnarray}
  \end{subequations}
\end{cor}
\begin{IEEEproof}
  Use \eqref{eq:upperboundonDeltawithbetaonly} and the fact $ 
  \sqrt{{\beta^{(1)}}}+ \sum_{i=1}^k\sqrt{{\beta^{(i)}}}\geq 
  (k+1)\sqrt{\beta_\tmin}$.
\end{IEEEproof}

The assumption \eqref{eq:betaminpositive} that the step-size sequence is 
lower-bounded by a strictly positive quantity is weaker than Lipschitz 
continuity of $\nabla \cL(\bx)$ because it is guaranteed to have 
$\beta_\tmin>\xi/L$ if $\nabla\cL(\bx)$ has a Lipschitz constant $L$. 


According to Corollary~\ref{thm:conv}, the \gls{PNPG} iteration attains 
$\mathcal{O}(k^{-2})$ convergence rate as long as
the cumulative error term \eqref{eq:E}
converges:
\begin{IEEEeqnarray}{c}
  \label{eq:Econverges}
  \mathcal{E}^{(+\infty)} \df \lim_{k\rightarrow+\infty} 
  \mathcal{E}^{(k)}<+\infty
\end{IEEEeqnarray}
which requires that $\theta^{(k)}\varepsilon^{(k)}$ decreases at a rate of 
$\mathcal{O}\PARENSbig{k^{-q}}$ with $q>0.5$.
This condition, also key for establishing convergence of iterates in 
Theorem~\ref{thm:convItr}, motivates us to use decreasing convergence 
criteria \eqref{eq:innerCrit} for the inner proximal-mapping iterations.
If the \gls{PG} steps in our \gls{PNPG} iteration are \emph{exact}, then 
$\varepsilon^{(i)}=0$ for all $i$, thus  $\mathcal{E}^{(k)}=0$ for all $k$, 
and the bound in \eqref{eq:objRateBetaMin} clearly ensures
$\mathcal{O}(k^{-2})$ convergence rate.

We now contrast our result in Theorem~\ref{thm:conv} with existing work on 
accommodating inexact proximal mappings in \gls{PG} schemes.  By 
recursively generating a function sequence that approximates the objective 
function, \cite{Villa2013Inexact} gives an asymptotic analysis of the 
effect of $\varepsilon^{(i)}$ on the convergence rate of accelerated 
\gls{PG} methods with inexact proximal mapping.  However, no explicit upper 
bound is provided for $\Delta^{(k)}$.  \citeauthor{Schmidt2011Inexact}
\cite{Schmidt2011Inexact} provide convergence-rate analysis and upper bound 
on $\Delta^{(k)}$ but this analysis does not apply here because it relies 
on fixed step-size assumption, uses different form of acceleration
\cite[Prop.~2]{Schmidt2011Inexact},
and has no convex-set constraint.
\citeauthor{BonettiniLorisPortaPrato2015} 
\cite{BonettiniLorisPortaPrato2015} also provide analysis of the 
inexactness of proximal mapping but for \gls{PG} methods without 
acceleration. The step size is decreasing from 1 for each iteration.  Note 
that \cite{BonettiniLorisPortaPrato2015} is able to catch the local 
curvature by using a scaling matrix in addition to its Armijo-style line 
search.  \cite{AujolDossal2015} provides analysis on both the convergence 
of the objective function and the iterates with inexact proximal operator 
for the cases with $B^{(1)}=1$ and $\mathbb{n}=\infty$, i.e., with 
decreasing step size only; see also Section~\ref{sec:FISTA} for the 
connection to \gls{FISTA}.
In the following, we introduce the convergence of iterates analysis with 
adaptive step size:

\begin{thm}[Convergence of Iterates]
  \label{thm:convItr}
  Assume that \begin{enumerate}[label=\arabic*)]
    \item \label{th1cond}
      the conditions of Theorem~\ref{thm:conv} hold,
    \item \label{Econverges}
$\mathcal{E}^{(+\infty)}$ exists: \eqref{eq:Econverges} holds,
    \item \label{convitergammabcond}
      the momentum tuning constants $(\gamma,b)$ satisfy
            \begin{IEEEeqnarray}{c"c}
        \label{eq:convitergammabcond}
        \gamma>2,& b\in\SBRs{0,1/\gamma^2},
      \end{IEEEeqnarray}

    \item \label{stepsizeseqcond}
      the step-size sequence $\PARENSs{\beta^{(i)}}_{i=1}^{+\infty}$ is 
      bounded within the range $\SBRs{\beta_\tmin, \beta_\tmax},\,
      (\beta_\tmin>0)$.

\end{enumerate}
Consider the \gls{PNPG} iteration without restart where 
\eqref{eq:proxgradstep} in Iteration~$i$ is replaced with the inexact 
\gls{PG} step in \eqref{eq:proxgradstepE}.  Then, the sequence of 
\gls{PNPG} iterates $\bx^{(i)}$ converges weakly to a minimizer of 
$f(\bx)$.
\end{thm}
\begin{IEEEproof}
  See Appendix~\ref{app:convItr}.
\end{IEEEproof}

Observe that Assumption~\ref{convitergammabcond} requires a narrower range 
of $(\gamma,b)$ than \eqref{eq:gammab}: indeed 
\eqref{eq:convitergammabcond} is a strict subset of \eqref{eq:gammab}.  The 
intuition is to leave a sufficient gap between the two sides of 
\eqref{eq:thetaCond} so that their difference becomes a quantity that is 
roughly proportional to the growth of $\theta^{(i)}$, which is important 
for proving the convergence of signal iterates 
\cite{Chambolle2015Convergence}.  Although the momentum term 
\eqref{eq:thetaupdateMod} with $\gamma=2$ is optimal in terms of minimizing 
the upper bound on the convergence rate (see Theorem~\ref{thm:conv}), it 
appears difficult or impossible to prove convergence of the signal iterates 
$\step{\bx}{i}$ for this choice of $\gamma$ 
\cite{Chambolle2015Convergence}, because the gap between the two sides of 
\eqref{eq:thetaCond} is upper-bounded by a constant.


The lemmas and proof of Theorem~\ref{thm:convItr} in 
Appendix~\ref{app:convItr} are developed for the step-size adaptation 
scheme in Section~\ref{sec:stepsize}: in particular, we use the adaptation 
parameter $\xi$ to constrain the rate of step-size increase, see 
\eqref{eq:stepsizeadaptation}. However, this rate can also be trivially 
constrained using $\beta_\tmin$ and $\beta_\tmax$ from 
Assumption~\ref{stepsizeseqcond} of Theorem~\ref{thm:convItr}, allowing us 
to establish convergence of iterates for general step-size selection rather 
than that in Section~\ref{sec:stepsize}.\footnote{As before, 
  $\step{\beta}{i}$ only needs to satisfy the majorization condition 
  \eqref{eq:majorCond}.}  We prefer to bound the rate of step-size increase 
  using $\xi$, which is more general than than the crude choice based on 
  Assumption~\ref{stepsizeseqcond}. 

\citeauthor{BonettiniPortaRuggiero2015} \cite{BonettiniPortaRuggiero2015} 
use a projected acceleration, with a scaling matrix instead of the adaptive 
step size to capture the local curvature. Inspired by 
\cite{Chambolle2015Convergence}, \cite{BonettiniPortaRuggiero2015} 
establishes convergence of iterates, but does not provide analysis of 
inaccurate proximal steps.  Both \cite{Chambolle2015Convergence} and 
\cite{BonettiniPortaRuggiero2015} require non-increasing step-size 
sequence.

\subsection{$\mathcal{O}(k^{-2})$ Convergence Acceleration Approaches}
\label{sec:Okminustwoaccelerationapproaches}

There exist a few variants of acceleration of the \gls{PG} method that 
achieve $\mathcal{O}(k^{-2})$ convergence rate 
\cite[Sec.~5.2]{Becker2011TFOCS}.  One competitor proposed by 
\citeauthor{Auslender2006AT} in \cite{Auslender2006AT} and restated in 
\cite{Becker2011TFOCS} where it was referred to as AT, replaces 
\eqref{eq:nesterov}--\eqref{eq:proxgradstep} with
\begin{subequations}
  \label{eq:AT}
  \begin{IEEEeqnarray}{rCl}
    \label{eq:ATtildebx}
    \wbar{\bx}^{(i)}&=&
    \PARENSBig{1-\frac{1}{\theta^{(i)}}}\bx^{(i-1)}
    +\frac{1}{\theta^{(i)}}\wtilde{\bx}^{(i-1)}
    \\
    \label{eq:proxgradstepAT}
    \wtilde{\bx}^{(i)} &=&
    \proxpbig{\theta^{(i)}\beta^{(i)} u r}{
      \wtilde{\bx}^{(i-1)} - \theta^{(i)}\beta^{(i)}\nabla\cL( 
      \wbar{\bx}^{(i)} )}
    \\
    \label{eq:ATbx}
    \bx^{(i)} &=&
    \PARENSBig{1-\frac{1}{\theta^{(i)}}}\bx^{(i-1)}
    +\frac{1}{\theta^{(i)}}\wtilde{\bx}^{(i)}
  \end{IEEEeqnarray}
\end{subequations}
where $\step{\theta}{i} = \theta_{2,1/4}^{(i)}$ in 
\eqref{eq:thetaupdateMod}.
Here, $\step{\beta}{i}$ in the \gls{TFOCS} implementation 
\cite{Becker2011TFOCS} is selected using the aggressive search with 
$\mathbb{n}=\mathbb{m}=0$.  All intermediate signals in 
\eqref{eq:ATtildebx}--\eqref{eq:ATbx} belong to $C$ and do not require 
projections onto $C$.  However, as $\theta^{(i)}$ increases with $i$, step 
\eqref{eq:proxgradstepAT} becomes unstable, especially when iterative 
solver is needed for its proximal operation.
To stabilize its convergence, AT relies on periodic restart by resetting 
$\theta^{(i)}$  using \eqref{eq:restart} \cite{Becker2011TFOCS}.  However, 
the period of restart is a tuning parameter that is not easy to select.  
For a linear Gaussian model, this period varies according to the  condition 
number of the sensing matrix $\Phi$ \cite{Becker2011TFOCS}, which is 
generally unavailable and not easy to compute in large-scale problems.  For 
other models, there are no guidelines how to select the restart period.


%
%

In Section~\ref{sec:NumEx}, we show that AT converges slowly compared with 
\gls{PNPG}, which justifies the use of projection onto $C$ in 
\eqref{eq:nesterov} and \eqref{eq:nesterov}--\eqref{eq:proxgradstep} 
instead of \eqref{eq:ATtildebx}--\eqref{eq:ATbx}.
\gls{PNPG} usually runs uninterrupted (without restart) over long stretches 
and benefits from Nesterov's acceleration within these stretches, which may 
explain its better convergence properties compared with AT. PNPG may also 
be less sensitive than AT to proximal-step inaccuracies; we have 
established convergence-rate bounds for \gls{PNPG} in the presence of such 
inaccuracies  (see \eqref{eq:objRate} and \eqref{eq:objRateBetaMin})
whereas AT does not yet have such guarantees.


%
%

\subsection{Relationship with FISTA}
\label{sec:FISTA}

The \gls{PNPG} method can be thought of as a generalized \gls{FISTA} 
\cite{Beck2009FISTA} that accomodates convex constraints, more general 
\glspl{NLL},\footnote{\gls{FISTA} has been developed for the linear 
Gaussian model in Section~\ref{sec:linGauss}.}  and (increasing) adaptive 
step size; thanks to this step-size adaptation, \gls{PNPG} \emph{does not} 
require Lipschitz continuity of the $\nabla\cL(\bx)$.  We need $B^{(i)}$ in 
\eqref{eq:B} to derive theoretical guarantee for convergence speed of the 
\gls{PNPG} iteration; see Section~\ref{sec:convergence_analysis}.  
In contrast with \gls{PNPG}, \gls{FISTA} requires Lipschitz continuity of 
$\nabla\cL(\bx)$ and has a non-increasing step size $\beta^{(i)}$, which 
allows for setting $B^{(i)}=1$ in \eqref{eq:thetaupdateMod} for all $i$ 
(see Appendix~\ref{app:FISTA}); upon setting $\gamma=2$ and $b=1/4$,  this 
choice yields the standard \gls{FISTA} (and Nesterov's \cite{Nesterov1983}) 
update.


\section{Numerical Examples}
\label{sec:NumEx}

We now evaluate our proposed algorithm by means of numerical simulations.  
We consider the nonnegative $C=\mathamsbb{R}_+^p$ in \eqref{eq:nonneg}.
\Gls{RSE} is adopted as the main metric to assess the performance of the 
compared algorithms:
\begin{IEEEeqnarray}{c}
  \text{RSE}=\frac{\|\what{\bx}-\bx_\text{true}\|_2^2}{\|\bx_\text{true}\|_2^2}
  \label{eq:rse}
\end{IEEEeqnarray}
where $\bx_\text{true}$ and $\what{\bx}$ are the true and reconstructed 
signals, respectively.

All iterative methods that we compare use the convergence criterion 
\eqref{eq:convcondouteriteration} with
\begin{IEEEeqnarray}{c}
  \label{eq:epsilon} \epsilon = \num{e-6}
\end{IEEEeqnarray}
and have the maximum number of iterations $I_\tmax=10^4$.

In the presented examples, \gls{PNPG} uses  momentum tuning constants
\begin{IEEEeqnarray}{c}
  \label{eq:gammabnumex}
  (\gamma,b)=(2,0)
\end{IEEEeqnarray}
and adaptive step-size parameters $(\mathbb{n}, \mathbb{m}) = (4,4)$ 
(unless specified otherwise), $\xi=0.8$, (initial) inner-iteration 
convergence constant $\eta=10^{-2}$, and maximum number of inner iterations 
$J_\tmax=100$.

We apply the AT method \eqref{eq:AT}  implemented in the \gls{TFOCS} 
package \cite{Becker2011TFOCS} with a periodic restart every 200 iterations 
(tuned for its best performance) and our proximal mapping in 
Section~\ref{sec:proximalmapping}.  Our inner convergence criteria 
\eqref{eq:innerCrit} cannot be implemented in the  \gls{TFOCS} package 
(i.e., it requires editing its code). Hence,
we select the proximal mapping that has a relative-error inner convergence 
criterion that corresponds to replacing the right-hand side of 
\eqref{eq:TVcrit} with $\epsilon' \norm{ \step{\balpha}{i,j}}_2$ and 
\eqref{eq:admmCrit} with $\epsilon' \norm{ \step{\bs}{i,j} }_2$.
This relative-error inner convergence criterion is easy to incorporate into 
the \gls{TFOCS} package \cite{Becker2011TFOCS} and is already used by the 
\gls{SPIRAL} package; see \cite{spiral}. Here, we select
\begin{IEEEeqnarray}{rCl}
  \label{eq:epsilonprime}
  \epsilon' = \num{e-5}
\end{IEEEeqnarray}
for both AT and \gls{SPIRAL}.

All the numerical examples were performed on a Linux workstation with an 
Intel(R) Xeon(R) CPU E31245 (\SI{3.30}{\giga\hertz}) and \SI{8}{\giga\byte} 
memory.  The operating system is Ubuntu 14.04 LTS (64-bit).  The Matlab 
implementation of the proposed algorithms and numerical examples is 
available \cite{pnpg}.

\begin{figure*}
  \centering
  \def\width{0.22}
  \def\height{105}
  \begin{subfigure}[t]{\width\textwidth}
    \centering
    \includegraphics[width=\textwidth]{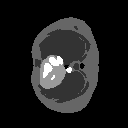}
    \caption{radio-isotope concentration}
    \label{fig:pet}
  \end{subfigure}
  \begin{subfigure}[t]{\width\textwidth}
    \centering
    \includegraphics[width=\textwidth]{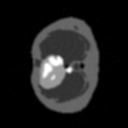}
    \putNumber{3.09}{\height} 
    \caption{FBP}
    \label{fig:fbp_pet}
  \end{subfigure}
  \begin{subfigure}[t]{\width\textwidth}
    \centering
    \includegraphics[width=\textwidth]{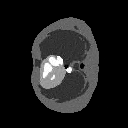}
    \putNumber{0.66}{\height}
    \caption{$\ell_1$} 
    \label{fig:pnpg_pet}
  \end{subfigure}
  \begin{subfigure}[t]{\width\textwidth}
    \centering
    \includegraphics[width=\textwidth]{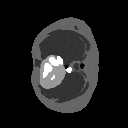}
    \putNumber{0.22}{\height}
    \caption{TV} 
    \label{fig:pnpg_TV_pet}
  \end{subfigure}
  \caption{
    (a) True emission image 
    and the reconstructions of the emission concentration map.
  }
  \label{fig:petExample}
\end{figure*}

\subsection{PET Image Reconstruction from Poisson Measurements}
\label{sec:poissonPET}

\begin{figure*}
  \centering
  \def\width{0.43}
  \begin{subfigure}[c]{\width\textwidth}
    \centering
    \makebox[0pt][c]{{\includegraphics{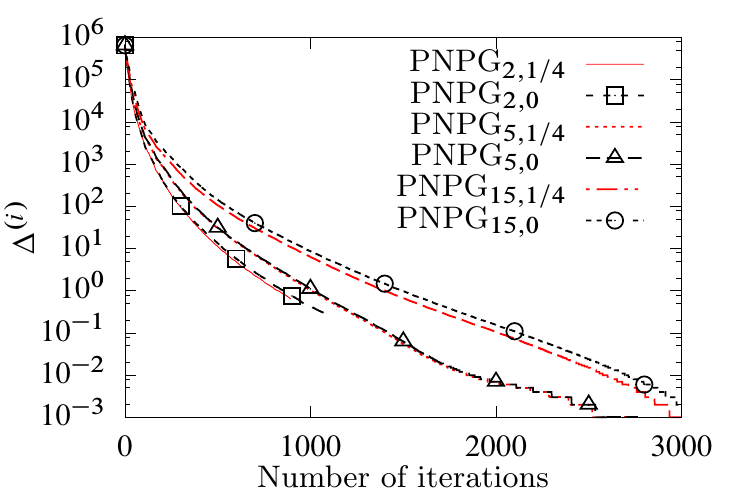}}}
    \caption{$\ell_1$, $\bm{1}^T\Phi\bx =10^8$, $a=0.5$}
    \label{fig:compareGammaB_l1}
  \end{subfigure}
  \begin{subfigure}[c]{\width\textwidth}
    \centering
    \makebox[0pt][c]{{\includegraphics{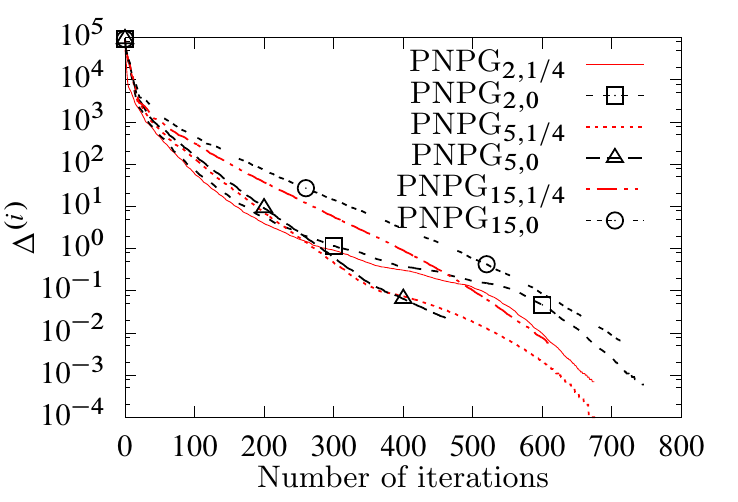}}}
    \caption{TV, $\bm{1}^T\Phi\bx =10^7$, $a=0$}
    \label{fig:compareGammaB_TV}
  \end{subfigure}
  \caption{
    Centered objectives of PNPG as functions of the \gls{CPU} time for (a) 
    $\ell_1$-norm and (b) \gls{TV} regularizations.
  }
  \label{fig:compareGammaB}
\end{figure*}

In this example, we adopt the Poisson \gls{GLM} \eqref{eq:poissonl} with 
identity link in \eqref{eq:identityLink}.  Consider \gls{PET} 
reconstruction of the $128\times128$ concentration map $\bx$ in 
Fig.~\ref{fig:pet}, which represents simulated radiotracer activity in the 
human chest.  Assume that the corresponding $128\times128$ attenuation map 
$\bkappa$ is known, 
which is needed to model the attenuation of the gamma rays 
\cite{OllingerFessler1997PET} and compute the sensing matrix $\Phi$ in this 
application.  We collect the photons from \num{90} equally spaced 
directions over \SI{180}{\degree}, with \num{128} radial samples in each 
direction.  Here, we adopt the parallel strip-integral matrix $\Gamma$ 
\cite[Ch.~25.2]{FesslerUnpublishedBook} and use its implementation in the 
\gls{IRT} \cite{irt} with sensing matrix
\begin{equation}
  \Phi= w \diag\bigl(\exp_\circ(-\Gamma\bkappa+\bc)\bigr)\Gamma
  \label{eq:petPhi}
\end{equation}
where $\bc$ is a known vector generated using a zero-mean \gls{iid} 
Gaussian sequence with variance $0.3$ to model the detector efficiency 
variation, and $w$ is a known scaling constant controlling the expected 
total number of detected photons due to true coincidence, 
$\bm{1}^T\Exp(\by-\bb) = \bm{1}^T\Phi\bx$, which is a \gls{SNR} measure.  
Here, we assume that the intercept term $\bb$ (generally nonzero) is due to 
the background radiation, scattering effect, and accidental coincidence 
combined together; see also \eqref{eq:identityLink} the Poisson \gls{GLM}.  
The elements of the intercept term have been set to a constant equal to 
\SI{10}{\percent} of the sample mean of $\Phi\bx$: 
$\bb=\frac{\bm{1}^T\Phi\bx}{10N}\bm{1}$.


The above model, choices of parameters in the \gls{PET} system setup, and 
concentration map have been adopted from  \gls{IRT} 
\cite[\texttt{emission/em\_test\_setup.m}]{irt}.  

Here, we consider both the $\ell_1$-norm with linear sparsifying transform 
[see \eqref{eq:Psibx}] and isotropic \gls{TV} penalties.  For 
\eqref{eq:Psibx}, we construct a $p \times p' =\num{12449 x 14056}$ 
sparsifying dictionary matrix $\Psi$ with orthonormal rows [which satisfies 
\eqref{eq:Psiorthogonalrows}] using the 2-D Haar \gls{DWT} with \num{6} 
decomposition levels and a full circular mask \cite{dgq11}.  

We compare the \gls{FBP} \cite{OllingerFessler1997PET} and \gls{PG} methods 
that aim at minimizing \eqref{eq:optgoal} with nonnegative $C$ in 
\eqref{eq:nonneg} and $\bpsi(\bx)$ in \eqref{eq:Psibx} and 
\eqref{eq:gradientMap}: PNPG-$\ell_1$, PNPG-TV, AT-$\ell_1$, AT-TV, and 
\gls{SPIRAL}-TV, where the suffixes ``-$\ell_1$'' and ``-TV'' denote the 
$\ell_1$-norm with linear sparsifying transform and \gls{TV} penalties.  We 
implemented \gls{SPIRAL}-TV using the centered \gls{NLL} term 
\eqref{eq:poissonl}, which improves the numerical stability compared with 
the original code in \cite{spiral}.  We do not compare with SPIRAL-$\ell_1$ 
because its inner iteration for the proximal step requires orthogonal (and 
hence square) $\Psi$, which is not the case here.




In this example, we adopt the following form of the regularization constant 
$u$:
\begin{IEEEeqnarray}{c}
  \label{eq:uPoissonidlink}
  u= 10^a.
\end{IEEEeqnarray}
We vary $a$ in \eqref{eq:uPoissonidlink} in the range $[-6,3]$ with a grid 
size of \num{0.5} and search for the reconstructions with the best average 
\gls{RSE} performance.  When possible, upper-bounding $u$ helps determine 
an appropriate search range for $u$; see \cite{NesterovTechReport} and 
\cite[Th.~\ref{asil-thm:uMAX}]{gdasil15} for such upper bounds for 
$\ell_1$-norm and \gls{TV} penalties, respectively.

All iterative methods were initialized by \gls{FBP} reconstructions 
implemented by \gls{IRT} \cite{irt}; see also 
\cite{OllingerFessler1997PET}.

Figs.~\ref{fig:fbp_pet}-\ref{fig:pnpg_TV_pet} show reconstructions for one 
random realization of the noise and detector variation $\bc$, with the 
expected total annihilation photon count (SNR) equal to \num{e8};  the 
optimal $a$ is 0.5.  At this SNR, all sparse reconstruction methods 
compared  (\gls{PNPG}, AT, and \gls{SPIRAL}) perform similarly as long as 
they employ the same penalty: the \gls{TV} sparsity penalty is superior to 
the $\ell_1$-norm counterpart;
%
see also Fig~\ref{fig:rmse_cntPET}.

\begin{figure*}
  \centering
  \def\width{0.43}
  \begin{subfigure}[c]{\width\textwidth}
    \centering
    \makebox[0pt][c]{{\includegraphics{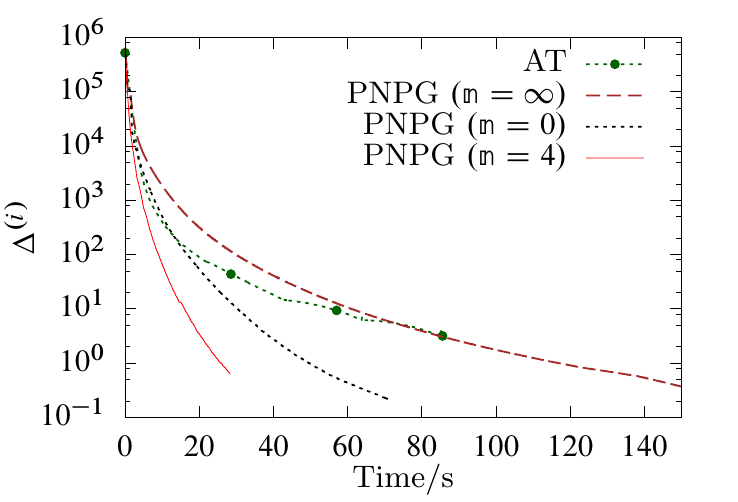}}}
    \caption{$\ell_1$, $\bm{1}^T\Phi\bx =10^8$, $a=0.5$}
    \label{fig:cost_timePET}
  \end{subfigure}
  \begin{subfigure}[c]{\width\textwidth}
    \centering
    \makebox[0pt][c]{{\includegraphics{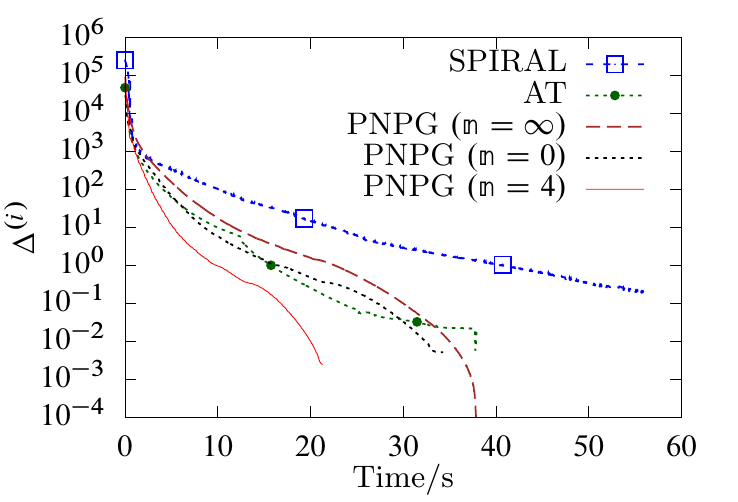}}}
    \caption{TV, $\bm{1}^T\Phi\bx  =10^7$, $a=0$}
    \label{fig:cost_timePET_TV}
  \end{subfigure}
  \caption{
    Centered objectives as functions of the \gls{CPU} time for (a) 
    $\ell_1$-norm and (b) \gls{TV} regularizations.
  }
  \label{fig:tracePET}
\end{figure*}

\begin{figure}
\def\width{0.43}
\centering
  \makebox[0pt][c]{{\includegraphics{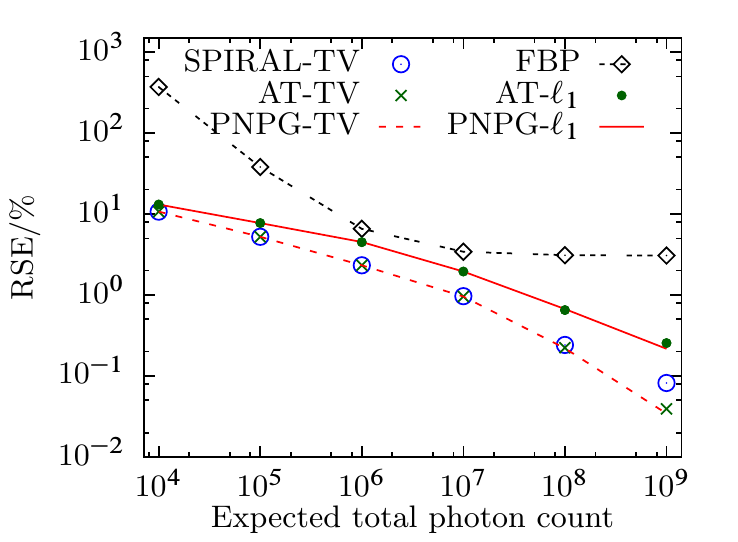}}}
\caption{Minimum average \glspl{RSE} as functions of $\bm{1}^T\Phi\bx$.}
\label{fig:rmse_cntPET}
\end{figure}

Figs.~\ref{fig:compareGammaB} and \ref{fig:tracePET} show the centered 
objectives $\Delta^{(i)}$ as  functions of \gls{CPU} time for the $\ell_1$- 
and \gls{TV}-norm signal sparsity regularizations and two random 
realizations of the noise and detector variation with different total 
expected photon counts.
%
Fig.~\ref{fig:compareGammaB} examines the convergence of PNPG as a function 
of the momentum tuning constants $(\gamma,b)$ in \eqref{eq:gammab}, using 
$\gamma \in \{2,5,15\}$ and $b \in \{0,1/4\}$.  For small $\gamma\leq5$, 
there is no significant difference between different selections and no 
choice is uniformly the best, consistent with 
\cite{Chambolle2015Convergence} which considers only $b=0$ and non-adaptive 
step size.  As we increase $\gamma$ further ($\gamma = 15$), we observe 
slower convergence.  
In the remainder of this section, we use $(\gamma,b)$ in 
\eqref{eq:gammabnumex}.
\looseness=-1


To illustrate the benefits of step-size adaptation, we present in 
Fig.~\ref{fig:tracePET} the performance of \PNPGinfty, which does not adapt 
to the local curvature of the \gls{NLL}, employs backtracking only, and has 
monotonically non-increasing step size, similar to \gls{FISTA}.  \PNPGf 
outperforms \PNPGinfty because it uses step-size adaptation; see also 
Fig.~\ref{fig:stepSizePoi} which corresponds to Fig.~\ref{fig:cost_timePET} 
and shows that the step size of \PNPGf for the $\ell_1$-norm signal 
sparsity penalty is consistently larger than that of \PNPGinfty.  The 
advantage of \PNPGf over the aggressive \PNPGz scheme is due to the 
\emph{patient} nature of its step-size adaptation, which leads to a better 
local majorization function of the \gls{NLL} and reduces time spent 
backtracking.  Indeed, if we do not account for the time spent on each 
iteration and only compare the objectives as functions of the iteration 
index, then \PNPGf and \PNPGz perform similarly; see 
\cite[Fig.~\ref{asil-fig:tracePET}]{gdasil15}.  Although \PNPGz and AT have 
the same step-size selection strategy and $\mathcal{O}(k^{-2})$ 
convergence-rate guarantees, \PNPGz converges faster; both schemes are 
further outperformed by \PNPGf.
Fig.~\ref{fig:cost_timePET_TV}
shows that \gls{SPIRAL}, which does not employ \gls{PG} step acceleration, 
is at least 3 times slower than \gls{PNPG}~$({\mathbb{n}=\mathbb{m}=4})$ 
for the same convergence threshold in \eqref{eq:epsilon}.

\begin{figure*}
\def\width{0.43}
\def\height{108}

\def\width{0.32}
\def\height{88}
\def\slen{8}
\centering
\begin{subfigure}[c]{\width\textwidth}
  \centering
  {\includegraphics{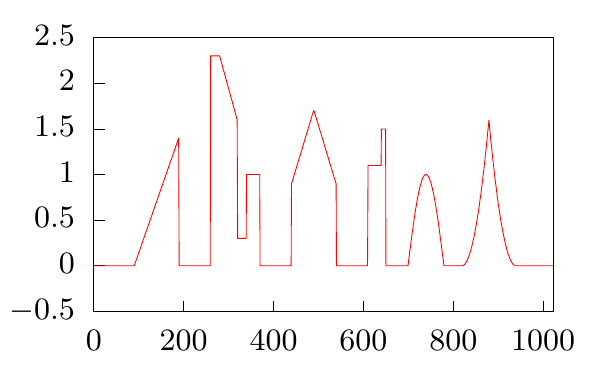}}
  \caption{true signal}
  \label{fig:skyline}
\end{subfigure}
{\begin{subfigure}[c]{\width\textwidth}
  \centering
  {\includegraphics{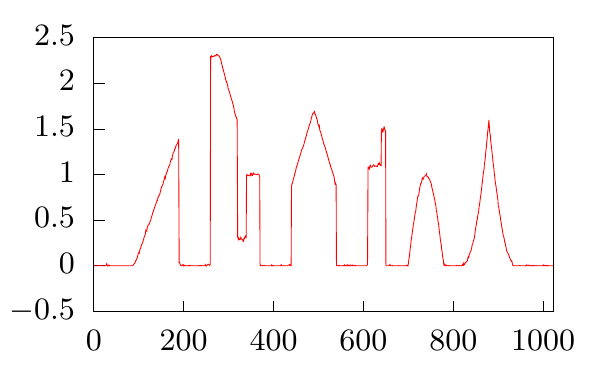}}
  \putNum{0.011}{\height}{\slen}{black}
  \caption{PNPG}
  \label{fig:recPNPG350}
\end{subfigure}}
{\begin{subfigure}[c]{\width\textwidth}
  \centering
  {\includegraphics{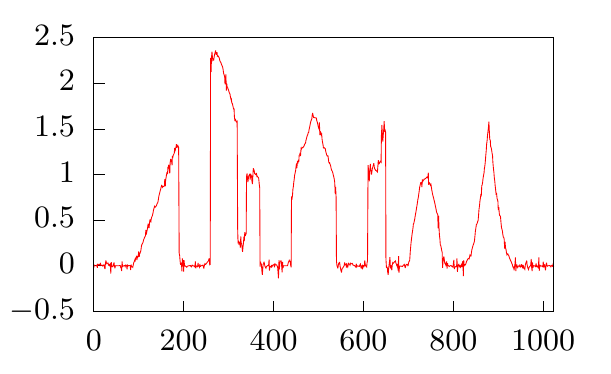}}
  \putNum{0.23}{\height}{\slen}{black}
  \caption{\gls{NPGS}}
  \label{fig:recFISTA350}
\end{subfigure}}
\caption{
  The true nonnegative skyline signal and its \gls{PNPG} and \gls{NPGS} 
  reconstructions for $N/p=0.34$.
}
\label{fig:skylinereconstructions}
\end{figure*}

Fig.~\ref{fig:rmse_cntPET} shows the minimum average (over 15 random 
realizations of the noise and detector variation $\bc$) \glspl{RSE} as 
functions of the expected total photon counts $\bm{1}^T\Phi\bx \in \CBRs{ 
10^4, 10^5, \ldots, 10^9}$, where $a$ has been selected to minimize the 
average \gls{RSE} for each method at each expected total photon count.  For 
$\ell_1$-norm and TV regularizations, the optimal $a$ increases from 
\num{-0.5} to \num{0.5} and \num{-0.5} to \num{1}, respectively, as we 
increase $\bm{1}^T\Phi\bx$ from $10^4$ to $10^{9}$.  As the \gls{SNR} 
increases, \gls{FBP} reaches a performance floor whereas \gls{PNPG}, AT, 
and \gls{SPIRAL} continue to improve thanks to the signal sparsity and 
nonnegativity constraints that they employ.  The \glspl{RSE} achieved by 
the methods that employ \gls{TV} regularization are 1.2 to 6.3 times 
smaller than those achieved by $\ell_1$-norm regularization.  As the 
\gls{SNR} increases, the convergence points of \gls{SPIRAL}-TV and 
\gls{PNPG}-TV diverge, which explains the difference between the 
\glspl{RSE} of the two methods at large \glspl{SNR} in 
Fig.~\ref{fig:rmse_cntPET}.  This trend is already observed when 
$\bm{1}^T\Exp(\by)=10^7$ in Fig.~\ref{fig:cost_timePET_TV}.\looseness=-1




\subsection{Skyline Signal Reconstruction from Linear Measurements}
\label{sec:linear1dex}

We adopt the $\ell_1$-norm penalty with $\bpsi(\cdot)$ in \eqref{eq:Psibx} 
and linear  measurement model with Gaussian noise in 
Section~\ref{sec:linGauss} where the elements of the sensing matrix $\Phi$ 
are \gls{iid}, drawn from the standard normal distribution.  Due to the 
widespread use of this measurement model, we can
compare wider range of methods than in the Poisson \gls{PET} example in 
Section~\ref{sec:poissonPET}.

We have designed a ``skyline'' signal of length $p=1024$ by overlapping 
magnified and shifted triangle, rectangle, sinusoid, and parabola 
functions; see Fig.~\ref{fig:skyline}.  We generate the noiseless 
measurements using $\by=\Phi\bx_\text{true}$.  The \gls{DWT} matrix $\Psi$ 
is constructed using the Daubechies-4 wavelet with 3 decomposition levels, 
whose approximation by the \SI{5}{\percent} largest-magnitude wavelet 
coefficients achieves $\RSE=\SI{98}{\percent}$.  
We compare: \begin{itemize}[leftmargin=0.15in,
  noitemsep,topsep=0pt,parsep=0pt,partopsep=0pt]
\item AT, \gls{PNPG}, and \gls{PNPG} with continuation 
  \cite{Wen2010FPCAS} (labeled \PNPGcont);
\item linearly constrained gradient projection
  method \cite{Harmany2010Gradient}, part of the \gls{SPIRAL} toolbox 
  \cite{spiral} and labeled \gls{SPIRAL} herein;
\item \gls{SpaRSA} \cite{Wright2009SpaRSA} with our implementation of the 
  proximal mapping in Section~\ref{sec:proximalmapping}, inner 
  convergence criterion with relative-error inner convergence criterion 
  and convergence threshold in \eqref{eq:epsilonprime}
  (easy to incorporate into the \gls{SpaRSA} software package 
  \cite{Wright2009SpaRSA} provided by the authors), and
  continuation (labeled \SpaRSAcont);
\item the \gls{GFB}  method \cite{Raguet2013GFB}:
  \begin{subequations}
    \label{eq:GFB}
    \begin{IEEEeqnarray}{rCl}
      \label{eq:z1}
      \bz_1 &\gets& \bz_1 + \lambda\SBRbig{
        \proxpbig{\frac{r}{w}u\norm{\Psi^T\cdot}_1}{2\bx-\bz_1-r\nabla\cL(\bx)} 
        -\bx
      }
\notag \\
      \\
      \bz_2 &\gets& \bz_2 + \lambda\SBRbig{
        \projpbig{C}{2\bx-\bz_2-r\nabla\cL(\bx)}-\bx
      }\\
      \bx &\gets& w\bz_1+(1-w)\bz_2
    \end{IEEEeqnarray}
  \end{subequations}
  with $r=1.8/\norm{\Phi}_2^2$, $\lambda=1$, and $w=0.5$ tuned for best 
  performance, and
\item the \gls{PDS} method \cite{Condat2013Primal}:    
  \begin{subequations}
    \label{eq:PDS}
    \begin{IEEEeqnarray}{rCl}
      \bar{\bz} &\gets&
      \projpbig{[-u,u]^p}{\bz+\sigma\Psi^T\bx}
      \\
      \bar{\bx} &\gets& 
      \projp{C}{\bx-\tau\nabla\cL(\bx)-\tau\Psi(2\bar{\bz}-\bz)}\\
      \bz &\gets& \bz + \rho(\bar{\bz}-\bz)\\
      \bx &\gets& \bx + \rho(\bar{\bx}-\bx)
    \end{IEEEeqnarray}
  \end{subequations}
  where $\tau$ and $\sigma$ are related as follows:
  $\tau(\sigma+\norm{\Phi}_2^2/2)=1$, with
  $\sigma=\tau$  and $\rho=2- 0.5 \norm{\Phi}_2^2 
  \PARENSbig{{1}/{\tau}-\sigma}^{-1}$ tuned for best performance,
\end{itemize}
all of which aim to solve the generalized analysis \gls{BPDN} problem with 
a convex signal constraint; the implementations of the \gls{GFB} and 
\gls{PDS} methods in  \eqref{eq:GFB} and \eqref{eq:PDS} correspond to this 
scenario.  Here, $p'=p$, $\Psi$ is an orthogonal matrix ($\Psi\Psi^T = 
\Psi^T \Psi = I$), and $\proxs{\lambda \norm{\Psi^T \cdot }_1 }{\ba} =
\Psi\softthrbig{\lambda}{\Psi^T\ba}$ has a closed-form solution
(see \eqref{eq:nonnegprojsoftthresh}),
which simplifies the implementation of the \gls{GFB} method (\eqref{eq:z1}, 
in particular); see the discussion in Section~\ref{sec:introduction}.
The other tuning options for \gls{SPIRAL}, \SpaRSAcont, and AT are kept to 
their default values, unless specified otherwise.

We initialize the iterative methods by the approximate minimum-norm 
estimate:
$\step{\bx}{0} = \Phi^T \SBRs{ \Exp(\Phi \Phi^T) }^{-1} \by
= \Phi^T \by/p$.
and select the regularization parameter $u$ as
\begin{IEEEeqnarray}{c"c}
u = 10^a U, & U \df \norm{\Psi^T\nabla\cL(\bm{0})}_\infty
\label{eq:u}
\end{IEEEeqnarray}
where $a$ is an integer selected from the interval $[-9,-1]$ and $U$ is an 
upper bound on $u$ of interest. Indeed, the minimum point $\bx^\star$ in 
\eqref{eq:xstar} reduces to \bm{0} if $u \geq U$ 
\cite[Sec.~\ref{report-sec:adaptivecontinuation}]{NesterovTechReport}.  


%

\begin{figure*}
\def\width{0.43}
\def\width{0.32}
\centering
\begin{subfigure}[c]{\width\textwidth}
  \centering
  \makebox[0pt][c]{{\includegraphics{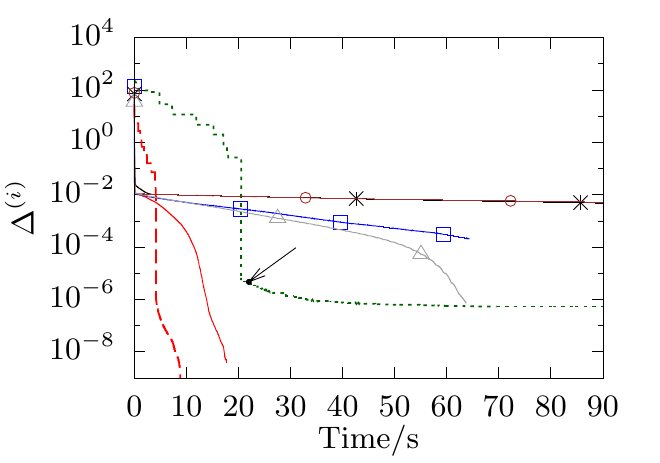}}}
  \caption{$a=-5, N/p=0.34$ }
  \label{fig:traceLinGaussNeg}
\end{subfigure}
\begin{subfigure}[c]{\width\textwidth}
  \centering
  \makebox[0pt][c]{{\includegraphics{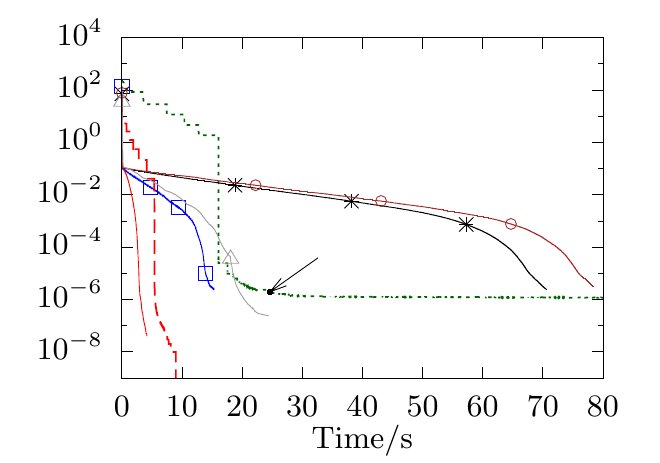}}}
  \caption{$a=-4, N/p=0.34$}
  \label{fig:traceLinGaussNeg3}
\end{subfigure}
\begin{subfigure}[c]{\width\textwidth}
  \centering
  \makebox[0pt][c]{{\includegraphics{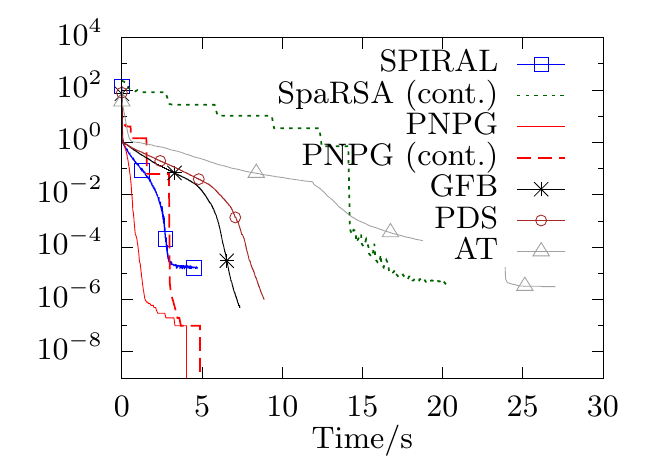}}}
  \caption{$a=-3,  N/p=0.34$}
  \label{fig:traceLinGaussNeg4}
\end{subfigure}
\begin{subfigure}[c]{\width\textwidth}
  \centering
  \makebox[0pt][c]{{\includegraphics{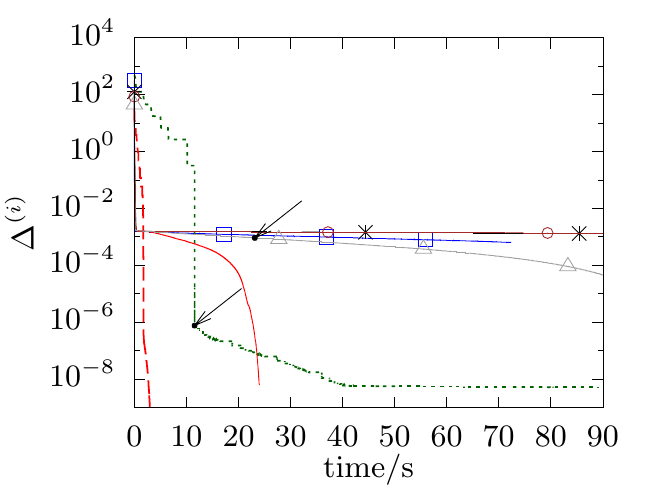}}}
  \caption{$a=-6, N/p=0.49$}
  \label{fig:traceLinGaussNegI}
\end{subfigure}
\begin{subfigure}[c]{\width\textwidth}
  \centering
  \makebox[0pt][c]{{\includegraphics{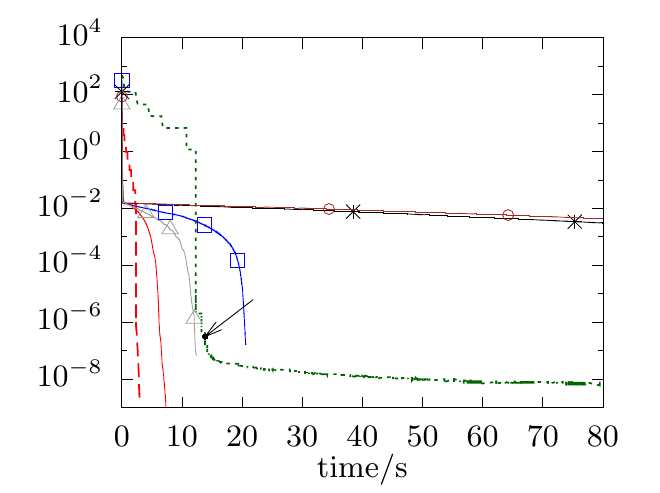}}}
  \caption{$a=-5, N/p=0.49$}
  \label{fig:traceLinGaussNeg3I}
\end{subfigure}
\begin{subfigure}[c]{\width\textwidth}
  \centering
  \makebox[0pt][c]{{\includegraphics{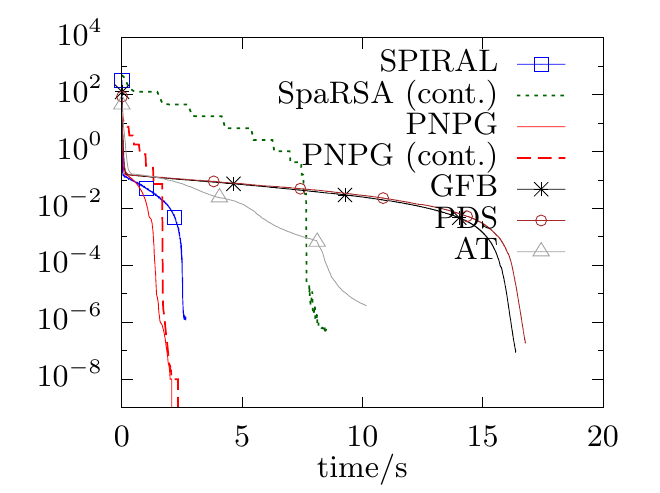}}}
  \caption{$a=-4,  N/p=0.49$}
  \label{fig:traceLinGaussNeg4I}
\end{subfigure}
\caption{
  Centered objectives as functions of CPU time for
  normalized numbers of measurements $N/p \in \{  0.34, 0.49 \}$ and 
  different regularization constants $a$.
  }
\label{fig:traceLinGauss}
\end{figure*}

As before, \PNPGf and \PNPGz converge at similar rates as functions of the 
number of iterations.  However, due to the excessive attempts to increase 
the step size at every iteration, \PNPGz spends more time backtracking and 
converges at a slower rate as a function of CPU time compared with \PNPGf; 
see also Fig.~\ref{fig:stepSizeLin} which corresponds to 
Fig.~\ref{fig:traceLinGaussNeg3} and shows the step sizes as functions of 
the number of iterations for $a=-4$ and $N/p=0.34$.  Hence, we present only 
the performances of \gls{PNPG} with $\mathbb{n}=\mathbb{m}=4$ in this 
section.


Fig.~\ref{fig:skylinereconstructions} shows the advantage brought by the 
convex-set nonnegativity signal constraints \eqref{eq:nonneg}.  
Figs.~\ref{fig:recPNPG350} and \ref{fig:recFISTA350} present the 
\gls{PNPG}~$(a=-5)$ and \gls{NPGS}~$(a=-4)$ reconstructions from one 
realization of the linear measurements with $N/p=0.34$ and $a$ tuned for 
the best \gls{RSE} performance.  Recall that \gls{NPGS} imposes signal 
sparsity only.  Here, imposing signal nonnegativity improves greatly the 
overall reconstruction and \emph{does not} simply rectify the signal values 
close to zero.

\begin{figure}
\def\width{0.43}
\centering
\makebox[0pt][c]{{\includegraphics[scale=1]{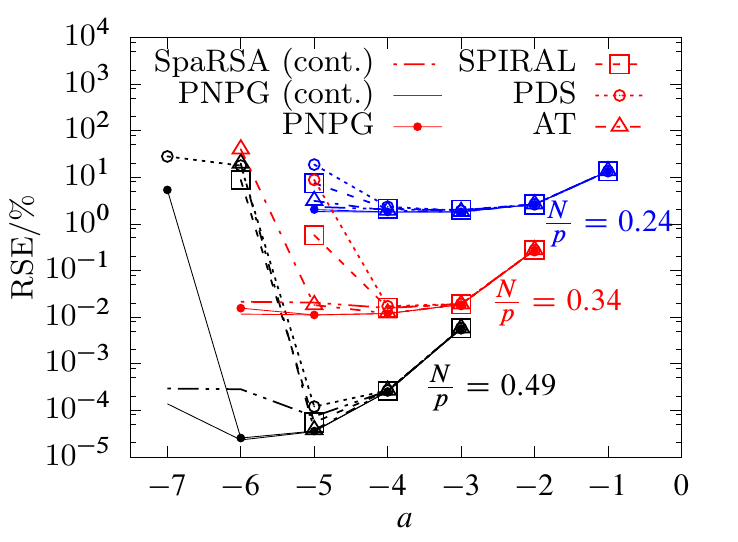}}}
\caption{ \kern-1pt Average \kern-1pt \glspl{RSE} \kern-1pt as \kern-1pt 
functions \kern-1pt of \kern-1pt  the \kern-1pt  regularization \kern-1pt 
constant
$a$.  
}
\label{fig:rmseVsA}
\end{figure}





Fig.~\ref{fig:traceLinGauss} presents the centered objectives 
$\Delta^{(i)}$ as functions of \gls{CPU} time for two random realizations 
of the sensing matrix $\Phi$ with different normalized numbers of 
measurements $N/p$ (in top and bottom rows, respectively) and several 
regularization constants $a$. The scenario with small $a$ is challenging 
for all optimization methods and the remedy is to use continuation. We 
present the performance of the \gls{PNPG} method both with and without 
continuation, labeled \gls{PNPG} and \PNPGcont, respectively.  \gls{SpaRSA} 
without continuation performs poorly and hence we apply only its version 
with continuation throughout this section.
We illustrate the benefits of continuation to the convergence of the 
\gls{PNPG} scheme when $a$ is small.  Note that the ``knee'' in the 
\SpaRSAcont performance curve occurs at the place where its continuation is 
completed, i.e., the regularization parameter for continuation descends to 
$u$.  This phenomenon is observed in all 20 trials.  Indeed, upon 
completion of continuation, \SpaRSAcont is simply a \gls{PG} scheme 
\emph{without acceleration}, which explains its low convergence rate 
following the ``knee''; we run \SpaRSAcont beyond its point of convergence 
mandated by \eqref{eq:epsilon}
and mark by arrows its convergence points for the convergence threshold in 
\eqref{eq:epsilon}.
Similarly, AT conveges prematurely in Fig.~\ref{fig:traceLinGaussNegI}, 
where its convergence point is marked by an arrow.

All methods in Fig.~\ref{fig:traceLinGauss} converge more slowly as $a$ 
decreases.   \gls{GFB}, \gls{PDS}, and \gls{SPIRAL} are especially 
vulnerable to small $a$; see Figs.~\ref{fig:traceLinGaussNeg} and 
\ref{fig:traceLinGaussNegI}.  Thanks to continuation, \SpaRSAcont and 
\PNPGcont have more stable \gls{CPU} times for different $a$s compared with 
the methods that do not employ continuation; \PNPGcont is slightly slower 
than \gls{PNPG} when $a$ is large.

Among the methods that do not employ continuation, \gls{PNPG} has the 
steepest descent rate, followed by \gls{SPIRAL} and {AT}.  AT uses 2 to 10 
times more \gls{CPU} time to reach the same objective than \gls{PNPG}.  
This justifies our convex-set projection in \eqref{eq:nesterov} for the 
Nesterov's acceleration step \cite{Nesterov1983,Beck2009FISTA}, shows 
superiority of \eqref{eq:nesterov} over AT's acceleration in 
\eqref{eq:ATtildebx} and \eqref{eq:ATbx}, and is consistent with the 
results in the Poisson example in Section~\ref{sec:poissonPET}.

In Fig.~\ref{fig:rmseVsA}, we show the average \glspl{RSE} (over 20 random 
realizations of the sensing matrix) as functions of the regularization 
parameter $a$ for normalized numbers of measurements $N/p \in \CBRs{0.24, 
0.34, 0.49}$, which are coded by blue, black, and red colors, respectively.  
\gls{PDS} and \gls{GFB} perform approximately the same, hence we show only 
the performance of \gls{PDS}.  \PNPGcont achieves the smallest \glspl{RSE} 
for all $a$, followed by \SpaRSAcont, where the gap between the two methods 
is due to the fact that \SpaRSAcont converges prematurely. 

\PNPGcont achieves the smallest \glspl{RSE} and is particularly effective 
for small $a$.  \gls{PNPG} struggles when $a$ is small, thus emphasizing 
the importance of continuation in this scenario.  \gls{SPIRAL} and 
\gls{PDS}  have similar performance and start to fail earlier than 
\gls{PNPG} as $a$ decreases and, for small $a$, yields reconstructions with 
much larger \glspl{RSE} than \gls{PNPG}.  
AT fails when $a<-5$ because it converges prematurely; see also 
Fig.~\ref{fig:traceLinGaussNegI}.

%

The methods that have large \gls{RSE} (around \SI{10}{\percent}) and 
effectively fail would not achieve better \gls{RSE} even if they use more 
stringent convergence criteria than \eqref{eq:epsilon}.

\section{Conclusion}
\label{sec:conclusion}

We developed a fast algorithm for reconstructing signals that are sparse in 
a transform domain and belong to a closed convex set by employing a 
projected proximal-gradient scheme with Nesterov's acceleration, restart and 
\emph{adaptive} step size. We applied the proposed framework to construct the 
first Nesterov-accelerated Poisson compressed-sensing reconstruction algorithm.
We presented integrated derivation of the proposed algorithm and 
convergence-rate upper-bound that accounts for inexactness of the proximal 
operator and also proved convergence of iterates.  Our \gls{PNPG} approach 
is computationally efficient compared with the state-of-the-art.




\appendices

\section{Derivation of Acceleration \eqref{eq:B}--\eqref{eq:nesterov}
and\\ Proofs of Lemma~\ref{thm:lll} and Theorem~\ref{thm:conv}}
\label{app:deracceleration}

\renewcommand{\theequation}{A\arabic{equation}}
\setcounter{equation}{0}

\renewcommand{\thesubsection}{\Alph{section}-\Roman{subsection}}
\renewcommand{\thesubsubsection}{\Alph{section}-\Roman{subsection}\alph{subsubsection}}
\renewcommand {\thesubsectiondis}{\Roman{subsection}}
\def\thesubsubsectiondis{\alph{subsubsection})}

We first prove Lemma~\ref{thm:lll} and then
derive the acceleration  \eqref{eq:B}--\eqref{eq:nesterov} and prove 
Theorem~\ref{thm:conv}.

\begin{IEEEproof}[Proof of Lemma~\ref{thm:lll}]
  According to Definition~\ref{eq:ePrecision} and  
  \eqref{eq:proxgradstepE},
\begin{subequations}
  \label{eq:lemma1Cond}
  \begin{IEEEeqnarray}{rCl}
    \label{eq:urGEQ}
    ur(\bx)
    &\geq&
    u r\PARENSs{\bx^{(i)}}
    + (\bx-\bx^{(i)})^T
    \biggl[
      \frac{
        \wbx^{(i)} -\bx^{(i)}
      }{\beta^{(i)}}
      -\nabla \cL( \wbx^{(i)})
    \biggr]
    \notag\\
    & &  \qquad - \: \frac{\PARENSs{\varepsilon^{(i)}}^2}{2\beta^{(i)}}
  \end{IEEEeqnarray}
  for any $\bx\in\mathamsbb{R}^p$.
  Moreover, due to the convexity of $\cL(\bx)$, we have
  \begin{IEEEeqnarray}{rCl}
    \label{eq:convexXB}
    \cL(\bx)&\geq&\cL\PARENSs{\wbx^{(i)}}
    +\PARENSs{\bx-\wbx^{(i)}}^T\nabla\cL\PARENSs{\wbx^{(i)}}.
  \end{IEEEeqnarray}
\end{subequations}
Summing \eqref{eq:urGEQ}, \eqref{eq:convexXB}, and \eqref{eq:majorCond} 
completes the proof.
\end{IEEEproof}

The following result from \cite[Proposition~2.2.1]{BertsekasOzdaglarNedic2003}
states that the distance between $\bx$ and $\by$ can be reduced by 
projecting them onto a closed convex set $C$.

\begin{lem}[Projection theorem]
\label{thm:convPrj}
The projection mapping onto a nonempty closed convex set $C\subseteq 
\mathamsbb{R}^p$ is nonexpansive
\begin{IEEEeqnarray}{c}
  \label{eq:projmapnonexpansive}
  \norm{\proj{C}{\bx}-\proj{C}{\by}}^2_2\leq\norm{\bx-\by}^2_2
\end{IEEEeqnarray}
for all $\bx,\by \in \mathamsbb{R}^p$.
\end{lem}

We now derive the Nesterov's acceleration step 
\eqref{eq:thetaupdateMod}--\eqref{eq:nesterov} with goal to select 
$\wbx^{(i)}$ in \eqref{eq:proxgradstep} that achieves the convergence rate 
of $\mathcal{O}(k^{-2})$.


Define sequences $\step{a}{i}>0$ and $\step{b}{i}>0$, multiply them with 
\eqref{eq:stari} and \eqref{eq:im1i}, respectively, add the resulting 
expressions, and multiply by $\step{\beta}{i}$ to obtain
\begin{IEEEeqnarray}{rCl}
\IEEEeqnarraymulticol{3}{l}{
  - 2\beta^{(i)}
  \step{c}{i}\Delta^{(i)}
  +2\beta^{(i)}
  \step{b}{i}\Delta^{(i-1)}
}
\nonumber
\\
&\geq&
\frac{1}{\step{c}{i}}
\normbig{
  \step{c}{i}\step{\bx}{i} -\step{b}{i}\step{\bx}{i-1} 
  -\step{a}{i}\bx^\star
}_2^2
\nonumber\\
& &  \qquad - \:  \frac{1}{\step{c}{i}}
\normbig{
  \step{c}{i}\wbx^{(i)} -\step{b}{i}\step{\bx}{i-1} -\step{a}{i}\bx^\star
}_2^2-c^{(i)}\PARENSs{\varepsilon^{(i)}}^2
\notag\\
&=&
\step{c}{i} \SBRbig{
  \step{t}{i}
  -   \step{\bar{t}}{i}  - \PARENSs{\varepsilon^{(i)}}^2
}
\label{eq:addedABineq}
\end{IEEEeqnarray}
where
\begin{subequations}
\begin{IEEEeqnarray}{rCl}
\label{eq:cdef}
\step{c}{i} &\df& \step{a}{i}+\step{b}{i}\\
\label{eq:titbari}
t^{(i)}&\df&\norm{
  \step{\bx}{i} - \bz^{(i)}
}_2^2, \qquad
\step{\bar{t}}{i} \df \norm{
  \step{\wbx}{i} -  \bz^{(i)}
 }_2^2
      \IEEEeqnarraynumspace
 \\
\label{eq:zinC}
\bz^{(i)} &\df& \frac{\step{b}{i}}{\step{c}{i}}\step{\bx}{i-1}
+\frac{\step{a}{i}}{\step{c}{i}} \bx^\star.
\end{IEEEeqnarray}
\end{subequations}
We arranged \eqref{eq:addedABineq} using completion of squares so that the 
first two summands are similar (but with opposite signs), with goal to 
facilitate cancellations as we sum over $i$.
Since we have control over the sequences $\step{a}{i}$ and $\step{b}{i}$, 
we impose the following conditions for $i\geq1$:
\begin{subequations}
\label{eq:goal}
\begin{IEEEeqnarray}{rCl}
  \label{eq:goal2}
  {\step{c}{i-1}}t^{(i-1)} &\geq& {\step{c}{i}}\bar{t}^{(i)}\\
  \label{eq:goal1}
  \pi^{(i)}&\geq&0
\end{IEEEeqnarray}
\end{subequations}
where
\begin{IEEEeqnarray}{rCl}
  \label{eq:piABC}
  \pi^{(i)}&\df&\beta^{(i)}c^{(i)}-\beta^{(i+1)}b^{(i+1)}.
\end{IEEEeqnarray}

Now, apply the inequality \eqref{eq:goal2} to the right-hand sides of 
\eqref{eq:addedABineq}:
\begin{subequations}
\begin{IEEEeqnarray}{rCl}
  \label{eq:addedABineqFinal}
  -{2\beta^{(i)}}\step{c}{i}\Delta^{(i)}
  +{2\beta^{(i)}}\step{b}{i}\Delta^{(i-1)}
  &\geq&
  {\step{c}{i}}t^{(i)}
  - {\step{c}{i-1}}t^{(i-1)}
  \nonumber\\
  & &  \kern5pt
  - \: c^{(i)}\PARENSs{\varepsilon^{(i)}}^2
\end{IEEEeqnarray}
and sum \eqref{eq:addedABineqFinal} over $i=1,2,\ldots,k$, which leads to 
summand cancellations and
\begin{IEEEeqnarray}{rCl}
  \IEEEeqnarraymulticol{3}{l}{
    -{2\beta^{(k)}}\step{c}{k}\Delta^{(k)}
    + 2\beta^{(1)} \step{b}{1}\Delta^{(0)}
    -2\sum_{i=1}^{k-1}\pi^{(i)}\Delta^{(i)}
  }
  \nonumber
  \\
  \label{eq:addedABineqSum0}
  & & \kern30pt \geq
  {\step{c}{k}}t^{(k)}
  -
  {\step{c}{0}}t^{(0)}
  -\sum_{i=1}^kc^{(i)}\PARENSs{\varepsilon^{(i)}}^2
  \\
  & & \kern30pt \geq
  -{\step{c}{0}}t^{(0)}
  -\sum_{i=1}^kc^{(i)}\PARENSs{\varepsilon^{(i)}}^2
  \label{eq:addedABineqSum}
\end{IEEEeqnarray}
 and \eqref{eq:addedABineqSum} follows from  \eqref{eq:addedABineqSum0}
by discarding the nonnegative term $\step{c}{k}t^{(k)}$.
\end{subequations}

Now, due to $\pi^{(i)}\Delta^{(i)}\geq0$, the inequality 
\eqref{eq:addedABineqSum} leads to
\begin{IEEEeqnarray}{rCl}
\label{eq:objRate2}
\Delta^{(k)} \leq
\frac{
  2\beta^{(1)}\step{b}{1}\Delta^{(0)}
  +{\step{c}{0}}t^{(0)}
  +\sum_{i=1}^kc^{(i)}\PARENSs{\varepsilon^{(i)}}^2
}{2\beta^{(k)}\step{c}{k}} \IEEEeqnarraynumspace
\end{IEEEeqnarray}
with simple upper bound on the right-hand side, thanks to summand 
cancellations facilitated by the assumptions \eqref{eq:goal}.

As long as $\step{\beta}{k}\step{c}{k}$ grows at a rate of $k^2$  and the 
inaccuracy of the proximal mappings leads to bounded  
$\sum_{i=1}^kc^{(i)}\PARENSs{\varepsilon^{(i)}}^2$, the centered objective 
function $\Delta^{(k)}$ can achieve the desired bound decrease rate of 
$1/k^2$.
Now, we discuss how to satisfy \eqref{eq:goal} and the growth rate of  
$\step{\beta}{k}\step{c}{k}$ by an appropriate selection of $\wbx^{(i)}$.

\subsection{Satisfying Conditions \eqref{eq:goal}}
\label{sec:goal0}

\subsubsection{Imposing equality in  \eqref{eq:goal2}}
\label{sec:xbareqxhat}
\eqref{eq:goal2} holds with equality for all $i$ and any $\bx^\star$ when 
we choose
$\wbx^{(i)}=\what{\bx}^{(i)}$ that satisfy
\begin{IEEEeqnarray}{rCl}
  \sqrt{\step{c}{i-1}}
  \PARENSs{
    \step{\bx}{i-1}
    -
    \step{\bz}{i-1}
  }
  =
  {\sqrt{\step{c}{i}}}
  \PARENSs{
    \what{\bx}^{(i)}
    - \step{\bz}{i}
  }.
  \label{eq:equal}
\end{IEEEeqnarray}
Now, \eqref{eq:equal} requires equal coefficients multiplying $\bx^\star$ 
on both sides, thus ${\step{a}{i}}/{\sqrt{\step{c}{i}}}=1/\mathrm{w}$ for 
all $i$, where $\mathrm{w}>0$ is a constant (not a function of $i$), which 
implies
$\step{c}{i} = \mathrm{w}^2\PARENSs{\step{a}{i}}^2$ and
$\step{b}{i}=\mathrm{w}^2\PARENSs{\step{a}{i}}^2-\step{a}{i}$,
see also \eqref{eq:cdef}.  Upon defining
\begin{subequations}
\label{eq:abc2theta}
\begin{IEEEeqnarray}{c}
  \label{eq:thetaDA}
  \step{\theta}{i} \df \mathrm{w}^2\step{a}{i}
\end{IEEEeqnarray}
we have
\begin{IEEEeqnarray}{c"c}
  \label{eq:abc2}
  \mathrm{w}^2\step{c}{i}=\PARENSs{\theta^{(i)}}^2,
  &
  \mathrm{w}^2\step{b}{i}=\PARENSs{\theta^{(i)}}^2-\theta^{(i)}.
\IEEEeqnarraynumspace
\end{IEEEeqnarray}
\end{subequations}
Plug \eqref{eq:abc2theta} into \eqref{eq:equal} and reorganize to obtain 
the following form of momentum acceleration:
\begin{IEEEeqnarray}{c}
\step{\what{\bx}}{i}=\step{\bx}{i-1}
+ \step{\Theta}{i} \PARENSs{\step{\bx}{i-1}-\step{\bx}{i-2}}.
\label{eq:momentum}
\end{IEEEeqnarray}

Although $\wbx^{(i)}=\step{\what{\bx}}{i}$ satisfies \eqref{eq:goal2}, it 
is not guaranteed to be within $\dom \cL$; consequently, the 
proximal-mapping step for this selection \emph{may not} be computable.  

\subsubsection{Selecting $\wbx^{(i)} \in C$ that satisfies
\eqref{eq:goal2}}
\label{app:selectingxbar}
We now seek $\wbx^{(i)}$ within $C$ that satisfies the inequality 
\eqref{eq:goal2}.  Since $\step{\bx}{i-1}$ and $\bx^\star$ are in $C$,
$\bz^{(i)} \in C$
by the convexity of $C$; see \eqref{eq:zinC}.
According to Lemma~\ref{thm:convPrj}, projecting \eqref{eq:momentum} onto 
$C$ preserves or reduces the distance between points.
Therefore,
\begin{IEEEeqnarray}{c}
\label{eq:wbx}
\wbx^{(i)} = P_C\PARENSs{\what{\bx}^{(i)}}
\end{IEEEeqnarray}
belongs to $C$ \emph{and} satisfies the condition \eqref{eq:goal2}:
\begin{subequations}
\begin{IEEEeqnarray}{rCl}
\label{eq:goal2ineq1}
  {\step{c}{i-1}}t^{(i-1)} &=& {\step{c}{i}}  \norm{
    \step{\what{\bx}}{i} -  \bz^{(i)}
 }_2^2\\
\label{eq:goal2ineq2}
 &\geq& {\step{c}{i}}  \norm{
    \step{\wbx}{i} -  \bz^{(i)}
 }_2^2  = {\step{c}{i}}\bar{t}^{(i)}
\end{IEEEeqnarray}
\end{subequations}
where \eqref{eq:goal2ineq1} and \eqref{eq:goal2ineq2}
 follow from \eqref{eq:equal} and by using Lemma~\ref{thm:convPrj}, 
 respectively; see also \eqref{eq:titbari}.

Without loss of generality, set $\mathrm{w}=1$ and rewrite and modify  
\eqref{eq:piABC},
\eqref{eq:titbari},  and \eqref{eq:addedABineqSum}  using 
\eqref{eq:abc2theta} to obtain
\begin{subequations}
\begin{IEEEeqnarray}{rCl}
  \pi^{(i)}
  &=&
  \beta^{(i)}\PARENSs{\step{\theta}{i}}^2
  \notag \\ & & - \: 
  \beta^{(i+1)}\step{\theta}{i+1}\PARENSbig{\theta^{(i+1)}-1},  \;\; i \geq 
  1
  \IEEEeqnarraynumspace
  \label{eq:pi}
  \\
  \label{eq:dueToabc}
\PARENSs{\step{\theta}{i}}^2{t}^{(i)}
  &=&
  \normbig{
    \theta^{(i)}\bx^{(i)}-\PARENSbig{\theta^{(i)}-1}\bx^{(i-1)}-\bx^\star
  }_2^2
  \IEEEeqnarraynumspace
  \\
    \sum_{i=1}^{k-1}\pi^{(i)}\Delta^{(i)}
  &\leq&
  \frac{1}{2}\SBRBig{
    \PARENSs{\step{\theta}{0}}^2t^{(0)}
    +\sum_{i=1}^k\PARENSs{\step{\theta}{i}\varepsilon^{(i)}}^2
  }
  \label{eq:addedABineqSumxx}
\end{IEEEeqnarray}
\end{subequations}
where \eqref{eq:addedABineqSumxx} is obtained by discarding the negative 
term $-2\beta^{(k)}\PARENSs{\theta^{(k)}}^2\Delta^{(k)}$ and the zero term 
$\beta^{(1)}\theta^{(1)}\PARENSs{\theta^{(1)}-1}\Delta^{(0)}$ (because 
$\theta^{(1)}=1$) on the left-hand side of \eqref{eq:addedABineqSum}.  Now, 
\eqref{eq:DueTo0Theta} follows from \eqref{eq:objRate2} by using 
$\step{\theta}{0}=\theta^{(1)}=1$ (see \eqref{eq:thetaupdateMod}), 
\eqref{eq:abc2theta}, and \eqref{eq:dueToabc} with $i=0$. 

\subsubsection{Satisfying \eqref{eq:goal1}}
\label{app:satgoal1}

By substituting \eqref{eq:pi} into \eqref{eq:goal1}, we obtain the 
conditions \eqref{eq:thetaCond} and interpret  
$\PARENSbig{\pi^{(i)}}_{i=1}^{+\infty}$ as the sequence of gaps between the 
two sides of \eqref{eq:thetaCond}.

\subsection{Connection to Convergence-Rate Analysis of \gls{FISTA}}
\label{app:FISTA}

If the step-size sequence $\PARENSs{\beta^{(i)}}$ is non-increasing (e.g., 
in the backtracking-only scenario with $\mathbb{n}=+\infty$),  
\eqref{eq:thetaupdateMod} with $B^{(i)}=1$ also satisfies the inequality 
\eqref{eq:solvedTheta}.  In this case, \eqref{eq:DueTo0Theta} still holds 
but \eqref{eq:upperboundonDeltawithbetaonly} does not because 
\eqref{eq:thetaGrow} no longer holds.  However, because $B^{(i)}=1$, we 
have $\theta^{(k)}\geq\frac{k+1}{2}$ and
\begin{IEEEeqnarray}{rCl}
\label{eq:fistaBound}
\Delta^{(k)}
&\leq&
\gamma^2\frac{
  \normbig{\step{\bx}{0}-\bx^\star}^2_2
  +\mathcal{E}^{(k)}
}{2\beta^{(k)}(k+1)^2}
\end{IEEEeqnarray}
which generalizes \cite[Th.~4.4]{Beck2009FISTA} to include the inexactness 
of the proximal operator and the convex-set projection.

\section{Convergence of Iterates}
\label{app:convItr}

\renewcommand{\theequation}{B\arabic{equation}}
\setcounter{equation}{0}


To prove convergence of iterates, we need to show that the centered 
objective function $\Delta^{(k)}$ decreases faster than the right-hand side 
of \eqref{eq:upperboundonDeltawithbetaonly}.
We introduce Lemmas~\ref{thm:deltaSummable} and \ref{thm:sumProdPi} and 
then use them to prove Theorem~\ref{thm:convItr}.
Throughout this Appendix, we assume that Assumption~\ref{th1cond} of 
Theorem~\ref{thm:convItr} holds, which justifies
\eqref{eq:twoineq} and \eqref{eq:thetaCond0} as well as results from 
Appendix~\ref{app:deracceleration} that we use in the proofs.

\begin{lem}
      \label{thm:deltaSummable}
      Under Assumptions~\ref{th1cond}--\ref{convitergammabcond} of 
      Theorem~\ref{thm:convItr},
\begin{IEEEeqnarray}{c}
      \label{eq:deltaSummable2}
      \sum_{i=1}^{+\infty}\PARENSbig{2\theta^{(i)}-1}\delta^{(i)}<+\infty.
    \end{IEEEeqnarray}
\end{lem}
\begin{IEEEproof}
By letting  $k\rightarrow+\infty$ in \eqref{eq:addedABineqSumxx} and using 
\eqref{eq:Econverges}, we obtain
\begin{IEEEeqnarray}{c}
      \label{eq:objSummable}
      \sum_{i=1}^{+\infty}
      \pi^{(i)}    \Delta^{(i)}<+\infty.
    \end{IEEEeqnarray}

For $i\geq1$, rewrite \eqref{eq:pi} using $\theta^{(i)}$ expressed in terms 
of $\theta^{(i+1)}$ (based on \eqref{eq:thetaupdateMod}):
\begin{IEEEeqnarray}{rCl}
  \label{eq:piGamma}
  {\pi^{(i)}}
  &=&
  \frac{\beta^{(i+1)}}{\gamma}\SBRBig{
    (\gamma-2)\theta^{(i+1)}+\frac{1-b\gamma^2}{\gamma}
  } \notag \\
  &\geq&
  \frac{\gamma-2}{\gamma}\beta^{(i+1)}\theta^{(i+1)}
\end{IEEEeqnarray}
where the inequality in \eqref{eq:piGamma} is due to $b\gamma^2-1<0$; see  
Assumption~\ref{convitergammabcond}.
Apply nonexpansiveness of the projection operator to \eqref{eq:im1i} and 
use  \eqref{eq:momentum} to obtain
\begin{IEEEeqnarray}{rCl}
  \label{eq:im1iDelta}
  {2\beta^{(i)}}
  \PARENSs{
    \step{\Delta}{i-1} -\step{\Delta}{i}
  }
  & \geq &
  \delta^{(i)}-\PARENSs{\Theta^{(i)}}^2\delta^{(i-1)}
  -\PARENSs{\varepsilon^{(i)}}^2
  \IEEEeqnarraynumspace
\end{IEEEeqnarray}
then multiply both sides of \eqref{eq:im1iDelta} by 
$\PARENSs{\theta^{(i)}}^2$, sum over $i=1,2,\ldots,k$ and reorganize:
\begin{subequations}
  \begin{IEEEeqnarray}{rCl}
    \IEEEeqnarraymulticol{3}{l}{
      \sum_{i=1}^{k-1}
      \PARENSs{2\theta^{(i)}-1}\delta^{(i)}
      \leq
      \PARENSs{\theta^{(0)}-1}^2\delta^{(0)}
      -\PARENSs{\theta^{(k)}}^2\delta^{(k)}
      +2\beta^{(1)}\Delta^{(0)}
    }
    \notag
    \\
    \label{eq:sumOfDelta}
    && +\: \mathcal{E}^{(k)}
    +
    2\sum_{i=1}^{k-1}
    \rho^{(i)}
    \Delta^{(i)}
    \\
    \label{eq:sumOfDeltaUpperBound}
    &\leq&
    2\beta^{(1)}\Delta^{(0)}
    +\mathcal{E}^{(k)}
    +\frac{4}{\gamma-2}\sum_{i=1}^{k-1}\pi^{(i)}\Delta^{(i)}
    \IEEEeqnarraynumspace
  \end{IEEEeqnarray}
  where (see \eqref{eq:pi})
\begin{IEEEeqnarray}{rCl}
  \label{eq:rho}
  \rho^{(i)}
  &=&
  \beta^{(i+1)}\PARENSs{\theta^{(i+1)}}^2
  -\beta^{(i)}\PARENSs{\theta^{(i)}}^2
  \\
  \label{eq:piIneq}
  &=&
  \beta^{(i+1)} \theta^{(i+1)} -  \pi^{(i)},
\end{IEEEeqnarray}
\end{subequations}
and we drop the zero term $\PARENSbig{\theta^{(0)}-1}^2\delta^{(0)}$ and 
the negative term $-\PARENSbig{\theta^{(k)}}^2\delta^{(k)}$ from 
\eqref{eq:sumOfDelta} and use the fact that $\rho^{(i)}\leq 
\tfrac{2}{\gamma-2}\pi^{(i)}$
implied by \eqref{eq:piGamma} to get \eqref{eq:sumOfDeltaUpperBound}.
Finally, let $k\rightarrow+\infty$ and use \eqref{eq:Econverges} and 
\eqref{eq:objSummable} to conclude \eqref{eq:deltaSummable2}.
\end{IEEEproof}

\begin{lem}
\label{thm:sumProdPi}
For $j \geq 3$,
\begin{IEEEeqnarray}{rCl}
    \label{eq:Pi_jj}
    \Pi_{j} \df\sum_{k=j}^{+\infty}\prod_{\ell=j}^k\Theta^{(\ell)}  &\leq& 
    \gamma\theta^{(j-1)}-1.
  \end{IEEEeqnarray}
\end{lem}
\begin{IEEEproof}
For $j \geq 3$,
  \begin{subequations}
  \begin{IEEEeqnarray}{rCl}
    \label{eq:difBetaTheta0}
    \frac{1}{\sqrt{\beta^{(k-1)}}\theta^{(k-1)}\theta^{(k)}}
    &\leq&
    \frac{\gamma}{\sqrt{\beta^{(k-1)}}\theta^{(k-1)}}
    -\frac{\gamma}{\sqrt{\beta^{(k)}}\theta^{(k)}}
    \qquad
    \\
    \label{eq:difBetaTheta}
    &\leq&
    \frac{\gamma}{\sqrt{\beta^{(k-2)}}\theta^{(k-2)}}
    -\frac{\gamma}{\sqrt{\beta^{(k)}}\theta^{(k)}}
  \end{IEEEeqnarray}
\end{subequations}
where we obtain the inequality \eqref{eq:difBetaTheta0} by combining the 
terms on the right-hand size and using \eqref{eq:grb1} and 
\eqref{eq:difBetaTheta} holds because $\sqrt{\beta^{(k)}}\theta^{(k)}$ is 
an increasing sequence (see Section~\ref{sec:convergence_analysis}).  
 Now,
\begin{subequations}
  \begin{IEEEeqnarray}{rCl}
    \label{eq:sumBetaTheta}
    \Pi_{j}
    &\leq&
    \sum_{k=j}^{+\infty}\prod_{\ell=j}^k
    \frac{\beta^{(\ell-2)}\PARENSbig{\theta^{(\ell-2)}}^2}{\beta^{(\ell-1)}\theta^{(\ell-1)}\theta^{(\ell)}}
    \label{eq:sumBetaTheta2}
    =
    \sum_{k=j}^{+\infty}
    \frac{\beta^{(j-2)}\PARENSbig{\theta^{(j-2)}}^2\theta^{(j-1)}}
    {\beta^{(k-1)}\PARENSbig{\theta^{(k-1)}}^2\theta^{(k)}} \notag
    \\
    \\
    \IEEEeqnarraymulticol{3}{l}{
      \leq
      \frac{\gamma\beta^{(j-2)}\PARENSbig{\theta^{(j-2)}}^2\theta^{(j-1)}}
      {\sqrt{\beta^{(j-2)}}\theta^{(j-2)}\sqrt{\beta^{(j-1)}}\theta^{(j-1)}}
      =
      \gamma \sqrt{B^{(j-1)}}
      \theta^{(j-2)}
    }
    \label{eq:Piijineq}
  \end{IEEEeqnarray}
\end{subequations}
where \eqref{eq:sumBetaTheta} follows by using \eqref{eq:Theta}, 
\eqref{eq:thetaCond} with $i = \ell-1$, and fraction-term cancellation;  
\eqref{eq:Piijineq} is obtained by substituting \eqref{eq:difBetaTheta} 
into \eqref{eq:sumBetaTheta} and canceling summation terms.  
\eqref{eq:Piijineq} implies \eqref{eq:Pi_jj} by using \eqref{eq:grb1} with 
$k=j-1$.
\end{IEEEproof}


Define
\begin{IEEEeqnarray}{c"c}
  \lambda^{(i)}\df \norm{\bx^{(i)}-\bx^\star}_2^2, &\Lambda^{(i)} \df 
  \lambda^{(i)}-\lambda^{(i-1)}.
\end{IEEEeqnarray}
Since $f\PARENSs{\bx^{(i)}}$ converges to $f(\bx^\star)=\min_\bx f(\bx)$ as 
the iteration index $i$ grows
and $\bx^{\star}$ is a minimizer, it is sufficient to prove the convergence 
of $\lambda^{(i)}$, see \cite[Th.~4.1]{Chambolle2015Convergence}.

\begin{IEEEproof}[Proof of Theorem~\ref{thm:convItr}]
Use \eqref{eq:stari} and the fact that $\Delta^{(i)}\geq0$ to get
\begin{IEEEeqnarray}{rCl}
\label{eq:fromStarI}
0 & \geq &
\lambda^{(i)}-\norm{\wbx^{(i)}-\bx^\star}_2^2 
-\PARENSs{\varepsilon^{(i)}}^2.
\end{IEEEeqnarray}
Now,
\begin{subequations}
\begin{IEEEeqnarray}{rCl}
    \IEEEeqnarraymulticol{3}{l}{
  \norm{\wbx^{(i)}-\bx^\star}_2^2
\leq \norm{
  \step{\what{\bx}}{i}-\bx^{\star}}_2^2
  =\lambda^{(i-1)}+\PARENSs{\Theta^{(i)}}^2\delta^{(i-1)}}
\nonumber\\
&&\; + \: 
2\Theta^{(i)}\PARENSs{\bx^{(i-1)}-\bx^{\star}}^T\PARENSs{\bx^{(i-1)}-\bx^{(i-2)}}
\label{eq:appnonexpansive}
\\
&\leq&\lambda^{(i-1)}+\PARENSs{\Theta^{(i)}}^2\delta^{(i-1)}
+\Theta^{(i)}\PARENSs{\Lambda^{(i-1)}+\delta^{(i-1)}}
  \IEEEeqnarraynumspace
\label{eq:usedNonexpansive}
\end{IEEEeqnarray}
\end{subequations}
where \eqref{eq:appnonexpansive} and \eqref{eq:usedNonexpansive} follow by 
using the nonexpansiveness of the projection operator
(see also \eqref{eq:momentum})
and the identity
\begin{IEEEeqnarray}{rCl}
  2(\ba-\bb)^T(\ba-\bc)=\norm{\ba-\bb}^2_2+\norm{\ba-\bc}_2^2-\norm{\bb-\bc}_2^2
  \IEEEeqnarraynumspace
\end{IEEEeqnarray}
respectively.  Combine the inequalities \eqref{eq:usedNonexpansive} and 
\eqref{eq:fromStarI} to get
\begin{subequations}
  \begin{IEEEeqnarray}{rCl}
    \label{eq:fromStarII}
    \Lambda^{(i)}
    &\leq& \Theta^{(i)}
    \SBRbig{ \Lambda^{(i-1)} + \PARENSbig{\Theta^{(i)}+1}\delta^{(i-1)}}
 +
    \PARENSs{\varepsilon^{(i)}}^2
  \IEEEeqnarraynumspace
    \\
    & \leq &
    \Theta^{(i)}
    \PARENSbig{\Lambda^{(i-1)} + {2\delta^{(i-1)}}/{\xi}}
    +
    \PARENSs{\varepsilon^{(i)}}^2
    \label{eq:fromStarIII}
  \end{IEEEeqnarray}
\end{subequations}
where \eqref{eq:fromStarIII} is due to $1<\frac{1}{{\xi}}$ (see 
\eqref{eq:xi}) and the following
\begin{subequations}
\begin{IEEEeqnarray}{rCl}
  \Theta^{(i)}&<&\frac{\theta^{(i-1)}}{\theta^{(i)}}
  =
  \frac{\sqrt{\beta^{(i-1)}}\theta^{(i-1)}\sqrt{\beta^{(i)}}}
  {\sqrt{\beta^{(i)}}\theta^{(i)}\sqrt{\beta^{(i-1)}}}
  \\
  &<&
  \frac{\sqrt{\beta^{(i)}}}{\sqrt{\beta^{(i-1)}}}
  \leq
  \frac{1}{\sqrt{\xi}}
  <
  \frac{1}{{\xi}}
\end{IEEEeqnarray}
\end{subequations}
where we have used \eqref{eq:Theta} and that
$\sqrt{\beta^{(i)}}\theta^{(i)}$ is an increasing sequence,
$\beta^{(i)}/\beta^{(i-1)} \geq 1/\xi$ (see Section~\ref{sec:stepsize}), 
and \eqref{eq:xi}.

According to \eqref{eq:thetaGrow} and Assumption~\ref{stepsizeseqcond} that 
the sequence $(\beta^{(i)})$ is bounded, there exists an integer $J$ such 
that
\begin{IEEEeqnarray}{c"c}
  \theta^{(j-1)}\geq2, & \Theta^{(j)}\geq\frac{1}{\theta^{(j)}}>0
  \label{eq:thetaLargerThan}
\end{IEEEeqnarray}
for all $j\geq J$, where the second inequality follows from the first and 
the definition of  $\Theta^{(j)}$, see \eqref{eq:Theta}.
Then
\begin{subequations}
  \begin{IEEEeqnarray}{rCl}
    \label{eq:Omega}
    \Omega^{(i)}
    &\df&
    \max\PARENSs{0,\Lambda^{(i)}}
    \leq
    \Theta^{(i)}\SBRbigg{
      \Omega^{(i-1)}+\frac{2\delta^{(i-1)}}{\xi}
      +\frac{\PARENSs{\varepsilon^{(i)}}^2}{\Theta^{(i)}}
    }
    \notag
    \\
    \\
    \IEEEeqnarraymulticol{3}{l}{
      \leq
      \sum_{j=J}^i
      \SBRbigg{
        \frac{2\delta^{(j-1)}}{\xi}
        +\frac{\PARENSs{\varepsilon^{(j)}}^2}{\Theta^{(j)}}
      }
      \prod_{\ell=j}^i\Theta^{(\ell)}
      +
      \Omega^{(J-1)} \prod_{\ell=J}^i\Theta^{(\ell)}}
      \IEEEeqnarraynumspace
      \label{eq:OmegaUpper0}
    \end{IEEEeqnarray}
  \end{subequations}
for $i\geq J$, where the inequality in \eqref{eq:Omega} follows by 
combining the inequalities \eqref{eq:fromStarIII} and $\Omega^{(i-1)} \geq 
\Lambda^{(i-1)}$ and \eqref{eq:OmegaUpper0} follows by recursively applying 
inequality \eqref{eq:Omega} with $i$ replace by $i-1, i-2, \ldots, J$.
Now, sum the inequalities \eqref{eq:OmegaUpper0} over $i=J, J+1, \ldots, 
+\infty$ and exchange the order of summation over $i$ and $j$ on the 
right-hand side:
\begin{IEEEeqnarray}{rCl}
  \sum_{i=J}^{+\infty}\Omega^{(i)}
  &\leq&
  \sum_{j=J}^{+\infty}
  \Pi_{j}
  \SBRbigg{
    \frac{2\delta^{(j-1)}}{\xi}
    +\frac{\PARENSs{\varepsilon^{(j)}}^2}{\Theta^{(j)}}
  }
  +
  \Pi_{J} \Omega^{(J-1)}
  \IEEEeqnarraynumspace
  \label{eq:OmegaSum1}
\end{IEEEeqnarray}
where $\Pi_j$ is defined in Lemma~\ref{thm:sumProdPi}.

For $j\geq J\geq3$,
\begin{subequations}
  \label{eq:forBothTerm}
\begin{IEEEeqnarray}{rCl}
  \label{eq:for1stTerm}
  \gamma \, (2\theta^{(j-1)}-1)-\Pi_{j}
  &\geq&\gamma \, \PARENSs{\theta^{(j-1)}-1}+1>0
  \IEEEeqnarraynumspace
  \\
  \label{eq:for2ndTerm}
  2 \, \gamma \, (\theta^{(j-1)}-1)-\Pi_{j}
  &\geq&\gamma \, \PARENSs{\theta^{(j-1)}-2}+1>0
\end{IEEEeqnarray}
\end{subequations}
where the first and second inequalities in \eqref{eq:forBothTerm} follow by 
applying Lemma~\ref{thm:sumProdPi} and
\eqref{eq:thetaLargerThan}, respectively; consequently,
\begin{subequations}
\begin{IEEEeqnarray}{rCl}
  \label{eq:PIdeltaSum}
  \sum_{j=J}^{+\infty} \Pi_{j}\delta^{(j-1)}
  &\leq&
  \gamma\sum_{j=J}^{+\infty} \PARENSs{2\theta^{(j)}-1} \delta^{(j)}
  <+\infty
  \IEEEeqnarraynumspace
  \\
  \label{eq:Piepsilonthetasum}
 \sum_{j=J}^{+\infty}
  \Pi_{j}\frac{\PARENSs{\varepsilon^{(j)}}^2}{\Theta^{(j)}}
  &\leq&
  2\gamma
  \sum_{j=J}^{+\infty}
  \PARENSs{\varepsilon^{(j)}}^2\frac{\theta^{(j-1)}-1}{\Theta^{(j)}}
  \\
  \label{eq:due2Theta}
  &=&
  2\gamma
  \sum_{j=J}^{+\infty}
  \PARENSs{\varepsilon^{(j)}}^2{\theta^{(j)}}
  \\
  \label{eq:thetaESum}
  &\leq&
  2\gamma
  \sum_{j=J}^{+\infty}
  \PARENSs{\theta^{(j)}\varepsilon^{(j)}}^2
\end{IEEEeqnarray}
\end{subequations}
where  \eqref{eq:PIdeltaSum}
follows from \eqref{eq:for1stTerm} and Lemma~\ref{thm:deltaSummable}
(for the second inequality)
and \eqref{eq:Piepsilonthetasum}
follows by using
\eqref{eq:for2ndTerm}; \eqref{eq:due2Theta} and \eqref{eq:thetaESum} are 
due to \eqref{eq:Theta} and \eqref{eq:thetaLargerThan}, respectively.
Combine \eqref{eq:PIdeltaSum} and \eqref{eq:thetaESum} with 
\eqref{eq:OmegaSum1} to conclude that
\begin{IEEEeqnarray}{rCl}
  \label{eq:sumOmegafinite}
  \sum_{i=1}^{+\infty}\Omega^{(i)} &<&+\infty.
\end{IEEEeqnarray}

The remainder of the proof uses the technique employed by
\citeauthor{Chambolle2015Convergence}
to conclude the proof of
\cite[Th.~4.1 at p.~978]{Chambolle2015Convergence}, which we repeat for 
completeness.
Define $X^{(i)}\df\lambda^{(i)}-\sum_{j=1}^{i}\Omega^{(j)}$, which is 
lower-bounded
because $\lambda^{(i)}$ and $\sum_{j=1}^{i}\Omega^{(j)}$
are lower- and upper-bounded (by \eqref{eq:sumOmegafinite}), respectively.
Furthermore, $\PARENSbig{X^{(i)}}$ is an non-increasing sequence:
\begin{IEEEeqnarray}{rCl}
  X^{(i+1)}&=&\lambda^{(i+1)}-\Omega^{(i+1)}-\sum_{j=1}^{i}\Omega^{(j)}
  \leq X^{(i)},
\end{IEEEeqnarray}
where we used the fact that 
$\Omega^{(i+1)}\geq\Lambda^{(i+1)}=\lambda^{(i+1)}-\lambda^{(i)}$. Hence, 
$\PARENSbig{X^{(i)}}$ converges as $i\rightarrow+\infty$.  Since 
$\sum_{j=1}^{i}\Omega^{(j)}$ converges, $\PARENSs{\lambda^{(i)}}$ also 
converges.
\end{IEEEproof}



\printbibliography

\end{document}